\documentclass[pra, floatfix, reprint]{revtex4-2}
\usepackage[utf8]{inputenc}
\usepackage{natbib}

\usepackage{siunitx}
\usepackage{chemformula}
\usepackage{lipsum}

\usepackage{graphicx}
\usepackage{xcolor}

\usepackage{amsmath, amssymb}
\usepackage{dsfont}

\usepackage{hyperref}
\hypersetup{
    colorlinks=true,
    linkcolor=olive,
    citecolor=olive,
    urlcolor=blue,
}

\newcommand{\dd}{\mathrm d}
\newcommand{\ee}{\mathrm e}
\newcommand{\ii}{\mathrm i}

\newcommand{\boltzmann}{k_\mathrm B}

\newcommand{\dfdx}[2]{\frac{\dd #1}{\dd #2}}

\newcommand{\tr}{\operatorname{Tr}}

\renewcommand{\Im}[1]{\operatorname{Im}\left\{#1\right\}}
\renewcommand{\Re}[1]{\operatorname{Re}\left\{#1\right\}}
\newcommand{\expval}[1]{\left\langle#1\right\rangle}
\newcommand{\artanh}{\operatorname{artanh}}

\newcommand{\vb}[1]{\mathbf{#1}} 
 
\newcommand{\uv}[1]{\mathbf{e}_{#1}} 
\renewcommand{\tensor}[1]{\mathds{#1}} 
\newcommand{\ut}{\tensor{1}} 
\newcommand{\dpv}{\vb r_0} 

\newcommand{\drivingfreq}{\Omega_\mathrm{d}}
\newcommand{\drivinglw}{\Gamma_\mathrm{d}}
\newcommand{\Tenv}{T_\mathrm{env}}
\newcommand{\dresstensor}{\tensor T}

\newcommand{\SNR}{\mathrm{SNR}}
\newcommand{\fs}{\mathcal{S}} 
\newcommand{\nprad}{r}

\newcommand{\eqnref}[1]{Eq.~(\ref{#1})}
\newcommand{\figref}[1]{Fig.~\ref{fig#1}}
\newcommand{\subfigref}[2]{Fig.~\ref{fig#1}(#2)}

\begin{document}

\title{Interaction Between an Optically Levitated Nanoparticle and Its Thermal Image: Internal Thermometry via Displacement Sensing}

\author{Thomas \surname{Agrenius}}
\author{Carlos \surname{Gonzalez-Ballestero}}
\author{Patrick \surname{Maurer}}
\author{Oriol \surname{Romero-Isart}}
\affiliation{Institute for Quantum Optics and Quantum Information of the Austrian Academy of Sciences, A-6020 Innsbruck, Austria}
\affiliation{Institute for Theoretical Physics, University of Innsbruck, A-6020 Innsbruck, Austria}
\date{January 20, 2023. \href{https://journals.aps.org/prl/accepted/a1071Yf3K201657932a77372bad431da897709bec}{\textcolor{black}{\copyright  American Physical Society.}}}

\begin{abstract}
    We propose and theoretically analyze an experiment where displacement sensing of an optically levitated nanoparticle in front of a surface  can be used to measure the induced dipole-dipole interaction between the nanoparticle and its thermal image. This is achieved by using a surface that is transparent to the trapping light but reflective to infrared radiation, with a reflectivity that can be time modulated. This dipole-dipole interaction relies on the thermal radiation emitted by a  silica nanoparticle having sufficient temporal coherence to correlate the reflected radiation with the thermal fluctuations of the dipole. The resulting force is orders of magnitude stronger than the thermal gradient force and it strongly depends on the internal temperature of the nanoparticle for a particle-to-surface distance greater than two micrometers.  We argue that it is experimentally feasible to use displacement sensing of a levitated nanoparticle in front of a surface as an internal thermometer in ultrahigh vacuum. Experimental access to the internal physics of a levitated nanoparticle in vacuum is crucial to understand the limitations that decoherence poses to current efforts devoted to prepare a nanoparticle in a macroscopic quantum superposition state.
\end{abstract}

\maketitle

Today, it is experimentally possible to optically levitate a nanoparticle in vacuum~\cite{gonzalez-ballestero_Science_374:6564_2021} and (i)~feedback-cool its center-of-mass motion to the ground state~\cite{delic_Science_367:6480_2020, magrini_Nature_595:7867_2021, tebbenjohanns_Nature_595:7867_2021, kamba_Opt.Express_30:15_2022, ranfagni_Phys.Rev.Research_4:3_2022}, (ii)~place it near a surface~\cite{alda_Appl.Phys.Lett._109:16_2016, kuhn_Appl.Phys.Lett._111:25_2017, diehl_Phys.Rev.A_98:1_2018, winstone_Phys.Rev.A_98:5_2018, magrini_Optica_5:12_2018, shen_Optica_8:11_2021, montoya_Appl.Opt._61:12_2022}, (iii)~measure the induced dipole-dipole interaction with another nanoparticle levitated in a second optical tweezer~\cite{rieser_Science_377:6609_2022}, and (iv)~use displacement sensing to detect forces in the zeptonewton regime~\cite{li_NaturePhys_7:7_2011, gieseler_Phys.Rev.Lett._109:10_2012, gieseler_NaturePhys_9:12_2013, moore_Phys.Rev.Lett._113:25_2014, ranjit_Phys.Rev.A_93:5_2016, rider_Phys.Rev.Lett._117:10_2016, hempston_Appl.Phys.Lett._111:13_2017}. In this Letter, we propose to combine these experimental capabilities to measure the dipole-dipole interaction of an optically levitated nanoparticle with its {\em thermal} image (see \figref{1}(a)). This  interaction  depends on the internal temperature  of the nanoparticle for particle-to-surface distances comparable to the thermal wavelength. Hence, we propose to leverage displacement sensing for internal thermometry of a levitated nanoparticle in vacuum~\cite{millen_NatureNanotech_9:6_2014, hoang_NatCommun_7:1_2016, delord_Appl.Phys.Lett._111:1_2017, hebestreit_Phys.Rev.A_97:4_2018, riviere_AVSQuantumSci._4:3_2022}. 

\begin{figure}
    \centering
    \includegraphics{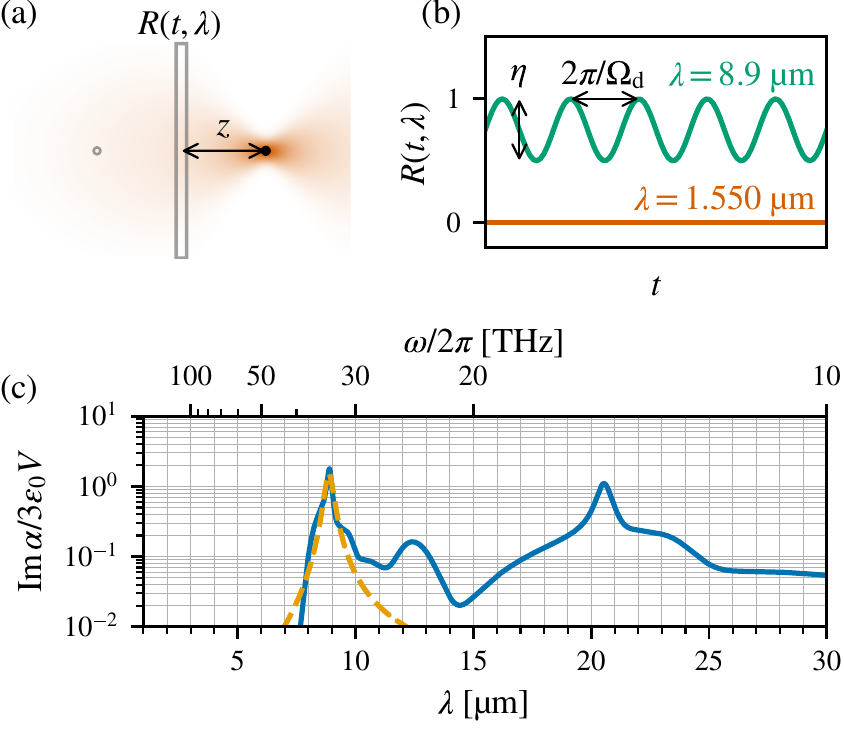}
    \caption{(a) Illustration of the proposed experiment. A nanoparticle is optically trapped at a distance $z$ from a surface with time- and wavelength-dependent reflection coefficient $R(t,\lambda)$. The effect of the surface can be approximated by the mirror image nanoparticle on the left. (b) The surface is transparent at the trapping wavelength, but the reflection coefficient can be modulated in time between high and low values at the peak wavelength of the nanoparticle's thermal emission. The modulation is sinusoidal with frequency $\drivingfreq$ and amplitude $\eta$. (c) The imaginary part of the polarizability of an \ch{SiO2} nanoparticle in the optical and infrared region of the spectrum using the model of $\epsilon(\lambda)$ from~\cite{kitamura_Appl.Opt._46:33_2007} (solid line). For $\lambda \in [1,7]\ \si{\micro m}$, the imaginary part of the polarizability is smaller than the lower limit of the graph~\cite{kitamura_Appl.Opt._46:33_2007}. At $\lambda=\SI{1.550}{\micro m}$ we use $\Im{\alpha}/(3\epsilon_0 V) = 1.5\times10^{-9}$ based on data reported in~\cite{bateman_Nat.Commun._5:1_2014, lee_NatCommun_3:1_2012}. The properties of the nanoparticle's infrared radiation is dominated by the peak at \SI{8.9}{\micro m}, to which we fit a Lorentzian function (dashed line).}
    \label{fig1}
\end{figure}

Our proposal aims not only at experimentally measuring an out-of-equilibrium Casimir force, which has been the object of intense research~\cite{obrecht_Phys.Rev.Lett._98:6_2007, henkel_J.Opt.A:PureAppl.Opt._4:5_2002, antezza_Phys.Rev.Lett._95:11_2005, antezza_Phys.Rev.Lett._97:22_2006, antezza_Phys.Rev.A_77:2_2008, kruger_EPL_2011, bimonte_Phys.Rev.A_92:3_2015, bimonte_Annu.Rev.Condens.MatterPhys._8:1_2017}, but also at giving experimental access to the internal physics of a levitated nanoparticle in vacuum which is very relevant for the field of levitodynamics~\cite{gonzalez-ballestero_Science_374:6564_2021,gonzalez-ballestero_Phys.Rev.Lett._124:9_2020}. Knowledge of the internal temperature $T$ of a nanoparticle and the imaginary part of its polarizability $\Im{\alpha}$ in the infrared regime is essential to quantify one of the most limiting sources of decoherence for experiments aiming to prepare a macroscopic quantum superposition of a nanoparticle ~\cite{romero-isart_Phys.Rev.Lett._107:2_2011,romero-isart_Phys.Rev.A_84:5_2011, bateman_Nat.Commun._5:1_2014, yin_Phys.Rev.A_88:3_2013, wan_Phys.Rev.Lett._117:14_2016, romero-isart_NewJ.Phys._19:12_2017, pino_QuantumSci.Technol._3:2_2018, stickler_NewJ.Phys._20:12_2018, weiss_Phys.Rev.Lett._127:2_2021, neumeier_FastQuantum_2022}: decoherence due to thermal emission of photons~\cite{romero-isart_Phys.Rev.A_84:5_2011,bateman_Nat.Commun._5:1_2014, hackermuller_Nature_427:6976_2004, schlosshauer_DecoherenceQuantumtoclassical_2007}. The associated decoherence rate approximately scales with $T^6$ and critically depends on $\Im{\alpha}$. Furthermore, the assignment of an internal temperature to the nanoparticle in out-of-equilibrium situations assumes that the nanoparticle internally equilibrates faster than any other relevant timescale. This assumption (known as the local equilibrium assumption) underlies the current understanding of the internal physics of levitated nanoparticles and the associated sources of decoherence, but whether it holds for a nanoparticle in ultrahigh vacuum or not is a question that needs to be answered experimentally~\cite{rubiolopez_Phys.Rev.B_98:15_2018}. Our experimental proposal would test this assumption and its consequences.

More specifically, we propose to optically trap a silica glass nanoparticle of radius $\nprad$ and mass $m$ at a  distance $z$ from a surface which is transparent at optical wavelengths but reflective at thermal (infrared) wavelengths, as illustrated in \subfigref{1}{a}. We assume that the interaction between the nanoparticle and the electromagnetic field of wavelengths $\lambda$ can be treated in the dipole approximation ($\nprad \ll \lambda$) with isotropic polarizability $\alpha(\lambda)$, which we calculate from bulk electric permittivity data $\epsilon(\lambda)$ for silica glass~\cite{kitamura_Appl.Opt._46:33_2007} using $\alpha(\lambda) = 3V\epsilon_0 [\epsilon(\lambda)-1][\epsilon(\lambda)+2]^{-1}$,
where $\epsilon_0$ is the vacuum permittivity and $V$ the volume of the nanoparticle. We display $\Im{\alpha(\lambda)}$ for infrared and optical wavelengths in \subfigref{1}{c}. To enable displacement sensing of the particle-surface interaction force, we propose to use a surface whose reflection coefficient can be modulated in time~\cite{kang_Adv.Opt.Mater._7:14_2019, huang_Rep.Prog.Phys._83:12_2020, yao_NanoLett._14:11_2014, zeng_LightSciAppl_7:1_2018, komar_Appl.Phys.Lett._110:7_2017}. As illustrated in \subfigref{1}{b}, we assume that the reflection coefficient~\footnote{We define the reflection coefficient $R(t,\lambda)$ as the magnitude of the Fresnel reflection coefficient at normal incidence.} can be modulated around a high value in the infrared, while simultaneously being practically zero and without modulation at the optical wavelength. As we discuss quantitatively later, high and unmodulated transparency at the optical wavelength is necessary to avoid the influence of an interaction force between the nanoparticle and its optical image in the force sensing experiment~\cite{rieser_Science_377:6609_2022, dania_Phys.Rev.Lett._129:1_2022, sm}. Similarly, since the charge on optically levitated nanoparticles can be controlled~\cite{frimmer_Phys.Rev.A_95:6_2017}, we propose to use electrically neutral particles to avoid interactions with the electrostatic mirror image~\cite{winstone_Phys.Rev.A_98:5_2018}. We note that all-optical cold damping schemes for electrically uncharged silica nanoparticles have recently been demonstrated~\cite{vijayan_ScalableAlloptical_2022, kamba_Opt.Express_30:15_2022}.

In order to evaluate the force acting on the particle in this scenario, we use the theory of fluctuational electrodynamics~\cite{henkel_J.Opt.A:PureAppl.Opt._4:5_2002, antezza_Phys.Rev.Lett._95:11_2005, antezza_Phys.Rev.Lett._97:22_2006, antezza_Phys.Rev.A_77:2_2008, kruger_EPL_2011, bimonte_Annu.Rev.Condens.MatterPhys._8:1_2017, buhmann_DispersionForcesI_2012}.
The steps we perform are summarized as follows (more details are given in~\cite{sm}): We treat the nanoparticle as an electric dipole $\vb d$ which has a part induced by the electric field $\vb d_\mathrm{ind}$ and a thermally fluctuating part $\vb d_\mathrm{th}$. The total electric field $\vb E$ is composed of a fluctuating field from the radiation of the nanoparticle $\vb E_\mathrm{ind}$, a second fluctuating field from thermal radiation from the environment $\vb E_\mathrm{th}$, and a non-fluctuating field due to the presence of the optical tweezer $\vb E_\mathrm{tw}$. The force acting on the nanoparticle is the expectation value of the standard force on an electric dipole in an electric field: $\vb F = \sum_{j\in\{x,y,z\}}\langle d_j(t) \nabla E_j(\vb r_0, t) \rangle$,
where $\vb r_0$ is the equilibrium position of the dipole.
To evaluate this expression, we diagonalize the total dipole moment $\vb d(t)$ and field $\vb E(\vb r_0, t)$ in terms of the input quantities: $\vb d_\mathrm{th}$ and $\vb E_\mathrm{th}$, whose correlations we assume to be given by the fluctuation--dissipation theorem, and $\vb E_\mathrm{tw}$. This is possible by transforming to the frequency domain where $\vb d_\mathrm{ind}(\omega) = \alpha(\omega)\vb E(\vb r_0, \omega)$ and $\vb E_\mathrm{ind}(\vb r, \omega) = \tensor G(\vb r, \vb r_0, \omega)\vb d(\omega)$, with  $\omega = 2\pi c/\lambda$, $c$ the speed of light in vacuum, and $\alpha(\omega)$ defined in terms of $\alpha(\lambda = 2\pi c/\omega)$ (see top axis of \subfigref{1}{c}). Here $\tensor G(\vb r, \vb r_0, \omega)$ is the electromagnetic Green's tensor. The presence of the surface modifies the Green's tensor, adding a scattering part $\tensor G_1$ to the free-space Green's tensor $\tensor G_0$ so that $\tensor G = \tensor G_0 + \tensor G_1$. After diagonalization, one finds
$\vb d(\omega) = \dresstensor (\dpv,\omega)\lbrace\vb d_\mathrm{th}(\omega) + \alpha(\omega)[\vb E_\mathrm{tw}(\vb r_0, \omega) + \vb E_\mathrm{th}(\vb r_0, \omega)]\rbrace$, and $\vb E(\vb r, \omega) = \vb E_\mathrm{tw}(\vb r, \omega) + \vb E_\mathrm{th}(\vb r, \omega) + \tensor G(\vb r, \vb r_0, \omega)\vb d(\omega)$. The tensor $\dresstensor (\dpv,\omega)\equiv[\tensor 1-\alpha(\omega)\tensor G_1(\vb r_0, \vb r_0, \omega)]^{-1}$ accounts for multiple reflections between the surface and the dipole. Since the subwavelength nanoparticle scatters radiation only weakly, reflections beyond the first order turn out to have a negligible impact on the forces and one can approximate $\dresstensor (\dpv,\omega)\approx \tensor 1$. $\vb F$ can now be evaluated with the fluctuation-dissipation relations $\expval{d_{j,\mathrm{th}}(\omega)d_{k,\mathrm{th}}(\omega')} = 2\pi\hbar \delta(\omega-\omega')[2n(\omega,T) + 1]\Im{\alpha(\omega)}\delta_{jk}$ and $\expval{E_{j,\mathrm{th}}(\vb r, \omega)E_{k,\mathrm{th}}(\vb r_0, \omega')} = 2\pi\hbar \delta(\omega-\omega')[2n(\omega,T_\mathrm{env}) + 1]\Im{G_{jk}(\vb r, \vb r_0, \omega)}$. Here $n(\omega, T)$ is the Bose--Einstein distribution, $\hbar$ is the reduced Planck's constant, and $T$ and $T_\mathrm{env}$ are the temperatures of the nanoparticle and the electromagnetic environment respectively. The quantities $\vb E_\mathrm{th}, \vb d_\mathrm{th}$, and $\vb E_\mathrm{tw}$ are assumed to have vanishing cross-correlations.

With this method, the total force $\vb F$ on the nanoparticle is found to have five contributions (see~\cite{sm} for details): (1) the optical force from the optical tweezer, (2) an interaction force between the nanoparticle and its optical mirror image \cite{dania_Phys.Rev.Lett._129:1_2022}, which we neglect in accord with our assumption that the surface is sufficiently transparent at the optical wavelength~\cite{sm}, (3) the zero-temperature Casimir force between the surface and the nanoparticle \cite{casimir_Phys.Rev._73:4_1948}, (4) a force due to interaction with environmental thermal radiation which depends on $\Tenv$, and (5) a force due to interaction between the nanoparticle and its reflected thermal radiation \cite{henkel_J.Opt.A:PureAppl.Opt._4:5_2002, kruger_EPL_2011}, or equivalently, its thermal mirror image. The last contribution depends on the nanoparticle internal temperature $T$, and we will denote it as
\begin{equation}
    \vb F_\mathrm{rad}(T,z) = \frac{\uv z \hbar c}{4\pi\epsilon_0 z^4}\int_0^\infty \frac{\dd\lambda}{\lambda^2}
    \frac{\Im{\alpha(\lambda)}f(2\pi z/\lambda)}{\exp{[2\pi\hbar c / (\boltzmann T \lambda)]} - 1}.
    \label{Frad}
\end{equation}
Here, $\uv z$ is the surface normal vector, $\boltzmann$ is Boltzmann's constant, and $f(x)$ is the oscillatory dimensionless function $f(x) = \Re{\exp{(2\ii x)}\left(-3 + 6\ii x + 6x^2 - 4\ii x^3\right)}$. The expressions of the other four contributions to $\vb F$ are given in~\cite{sm}.

\begin{figure}
    \centering
    \includegraphics{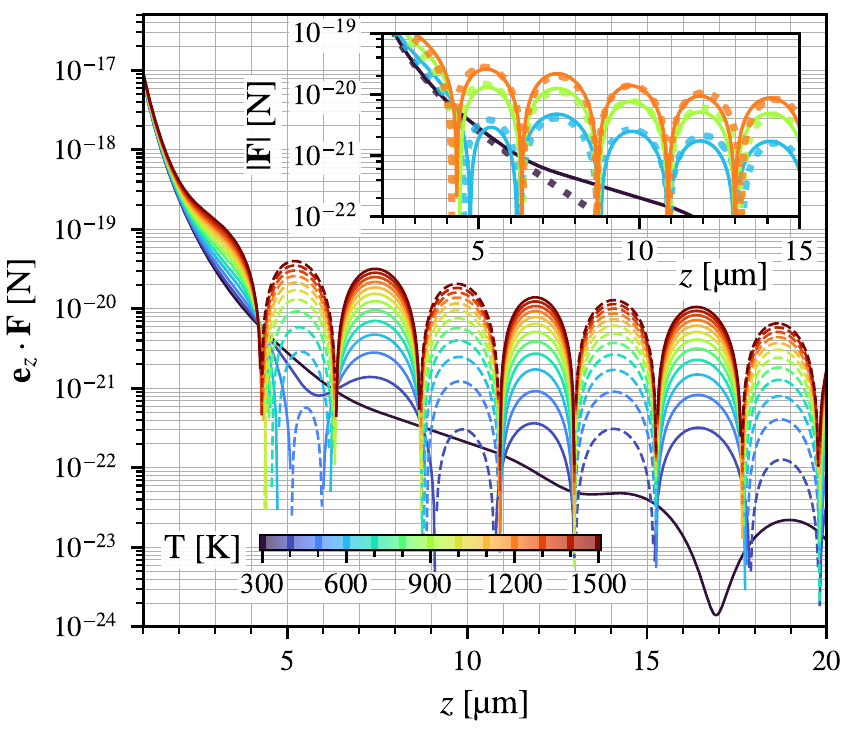}
    \caption{The total dipole force on the nanoparticle along the direction normal to the surface when $R(t,\lambda) = 1$ for $\lambda \geq \SI{7}{\micro m}$, using the full spectral dependence of $\alpha(\omega)$ (solid line in \subfigref{1}{c}). Solid/dashed lines indicate that the force is repulsive/attractive. The $T=\SI{300}{K}$  curve (black) is the force on the particle in equilibrium since we assume that $\Tenv=\SI{300}{K}$. Inset: Comparison of the total dipole force on the nanoparticle computed using the full spectral dependence (lines) versus the Lorentzian fit (dots) of $\alpha(\omega)$ as shown in \subfigref{1}{b}.  }
    \label{fig2}
\end{figure}

In \figref{2}, we plot  $\uv z \cdot \vb F$ as a function of $z$ and $T$ in the range 300 to \SI{1500}{K} with increments of \SI{100}{K}. We assume $\Tenv = \SI{300}{K}$. Note that therefore the black line ($T = \SI{300}{K}$)  corresponds to the equilibrium case. We find that as $T$ is increased, the total force becomes dominated by the $T$-dependent fifth contribution $\vb F_\mathrm{rad}$. Only at the smallest distances $z \lesssim  \SI{2}{\micro m}$ is the temperature dependence lost. This is due to the dominance of the zero-temperature Casimir force component over all the other force contributions at small separations~\cite{kruger_EPL_2011, buhmann_DispersionForcesI_2012, bimonte_Phys.Rev.A_92:3_2015}. At distances $z > \SI{2}{\micro m}$, the force scales more slowly with distance than the $z^{-5}$ scaling characteristic of zero-temperature Casimir--Polder forces~\cite{kruger_EPL_2011, casimir_Phys.Rev._73:4_1948}. Additionally, we observe that the force oscillates in sign along $z$ with a temperature-independent period.
The oscillations arise due to the peak at $\lambda_\mathrm{peak}=\SI{8.9}{\micro m}$ of $\Im{\alpha(\lambda)}$ (which is attributed to vibrations of the \ch{Si}--\ch{O} bond in silica glass~\cite{kitamura_Appl.Opt._46:33_2007}) dominating the integral in \eqnref{Frad} for $T\gtrsim\SI{400}{K}$. 
To confirm this explanation, we fit a Lorentzian function centered at $\lambda_\mathrm{peak}$ to $\Im{\alpha(\lambda)}$, finding that the fit has full width at half maximum (FWHM) of \SI{0.4}{\micro m} (dashed curve in \subfigref{1}{c}). In the inset of \figref{2}, we compare the total force calculated using the fitted and full $\Im{\alpha(\lambda)}$ (the two curves in \subfigref{1}{c}), finding excellent agreement when the temperature of the nanoparticle is above \SI{400}{K}. We remark that the oscillatory nature of the force could be utilized to perform a measurement of the distance from the nanoparticle to the surface.
 
We emphasize that the interaction giving rise to $\vb F_\mathrm{rad}$ is of temporally coherent nature~\cite{rieser_Science_377:6609_2022}. That is, the thermal dipole moment $\vb d_\mathrm{th}(t)$ remains correlated with itself during the time it takes for the thermally emitted radiation to be reflected by the surface and return to the particle: in time domain, $\vb E_\mathrm{ind}(t) = \int_{-\infty}^\infty \dd t'\, \tensor G(t-t')\vb d(t')  $, and therefore $\vb F_\mathrm{rad} \propto \expval{d_{\mathrm{th},j}(t)d_{\mathrm{th},k}(t')}$. The temporal coherence of the nanoparticle's thermal radiation is endowed by the narrow frequency spectrum of $\Im{\alpha}$. A hypothetical increasingly broadband emitter, with accordingly shorter coherence time, would produce a force  $\vb F_\mathrm{rad}$ with weakening oscillations which in the white-noise limit $\expval{d_{\mathrm{th},j}(t) d_{\mathrm{th},k}(t')} \to \delta(t-t')$ tends to zero since the time-domain Green's tensor vanishes for zero time argument. Additionally, we point out that $\vb F_\mathrm{rad}$ cannot be written as the gradient of an electromagnetic field intensity~\cite{sonnleitner_Phys.Rev.Lett._111:2_2013, haslinger_NaturePhys._14:3_2018}. The gradient force that the nanoparticle experiences due to its radiated field intensity is a higher-order term in the tensor $\dresstensor(\dpv,\omega)$ which is not included in \eqnref{Frad} and the contribution of which to $\vb F$ is negligible.

\begin{table}[b]
    \caption{Table of proposed experimental parameters.}
    \label{tab:parameters}
    \begin{tabular}{ll}
        Parameter & Value \\ \hline
        nanoparticle radius $\nprad$ &    \SI{100}{nm} \\
        nanoparticle density $\rho$ & \SI{2200}{kg\ m^{-3}} \\
        environment temperature $\Tenv$ & \SI{300}{K} \\
        gas pressure $p_*$ & \SI{e-10}{mbars} \\
        tweezer wavelength $\lambda_0$ & \SI{1.550}{\micro\meter} \\
        numerical aperture $N$ & 0.75 \\
        laser power $P_\mathrm{tw}$ & \SI{10}{mW} \\
        reflection modulation amplitude $\eta$ \hspace{5mm} & 0.5 \\
        driving frequency $\drivingfreq$ & $2\pi \times \SI{12}{kHz}$ \\
        driving quality factor $Q$ & $10^6$ \\
    \end{tabular}
\end{table}

\figref{2} shows that $\vb F$ reaches values above $\SI{e-21}{N}$, the currently demonstrated sensitivity in dynamic force sensing experiments with optically levitated nanoparticles in vacuum~\cite{geraci_Phys.Rev.Lett._105:10_2010, li_NaturePhys_7:7_2011, gieseler_Phys.Rev.Lett._109:10_2012, gieseler_NaturePhys_9:12_2013, ranjit_Phys.Rev.A_93:5_2016, jain_Phys.Rev.Lett._116:24_2016, hempston_Appl.Phys.Lett._111:13_2017,delic_Science_367:6480_2020, magrini_Nature_595:7867_2021, tebbenjohanns_Nature_595:7867_2021}, for a wide range of temperatures $T$ and particle-surface separations $z$. This suggests that $\vb F_\mathrm{rad}(T,z)$ can be measured with current laboratory capabilities.
The thermally limited force sensitivity $\fs$ is defined as $\fs = \sqrt{S_\mathrm{noise}}$, where $S_\mathrm{noise}$ is the power spectral density (PSD) of the thermal forces acting on the nanoparticle. Under cold damping of the nanoparticle motion, the leading thermal forces are impacts with residual gas molecules and photon shot noise~\cite{jain_Phys.Rev.Lett._116:24_2016}, and it can be shown that~\cite{sm}
\begin{equation}
    \fs = \sqrt{\frac{2\hbar P_\mathrm{scatt}}{5c\lambda_0} + \frac{m\gamma_\mathrm{gas}\boltzmann T_\mathrm{env}}{\pi}}. \label{force_sensitivity}
\end{equation}
Here $P_\mathrm{scatt}$ is the power that the nanoparticle scatters from the laser beam of wavelength $\lambda_0$~\cite{jain_Phys.Rev.Lett._116:24_2016}, and $\gamma_\mathrm{gas}$ is the damping rate due to collisions with gas molecules, directly proportional to the vacuum chamber pressure $p$~\cite{beresnev_MotionSpherical_1990}. We assume that the contributions from other sources of noise (e.g. surface-induced noise on neutral particles~\cite{gehm_Phys.Rev.A_58:5_1998}) are negligible compared to the photon shot noise. In \subfigref{3}{a} we show that for the choice of experimental parameters presented in Table~\ref{tab:parameters}, zeptonewton force sensitivity is achieved for $p \leq \SI{e-9}{\milli\bar}$ with saturation to the photon shot noise limit \SI{4e-22}{N/\sqrt{Hz}} at $p_* = \SI{e-10}{\milli\bar}$. We assume this value of the pressure for the remainder of the discussion.

\begin{figure}
    \includegraphics{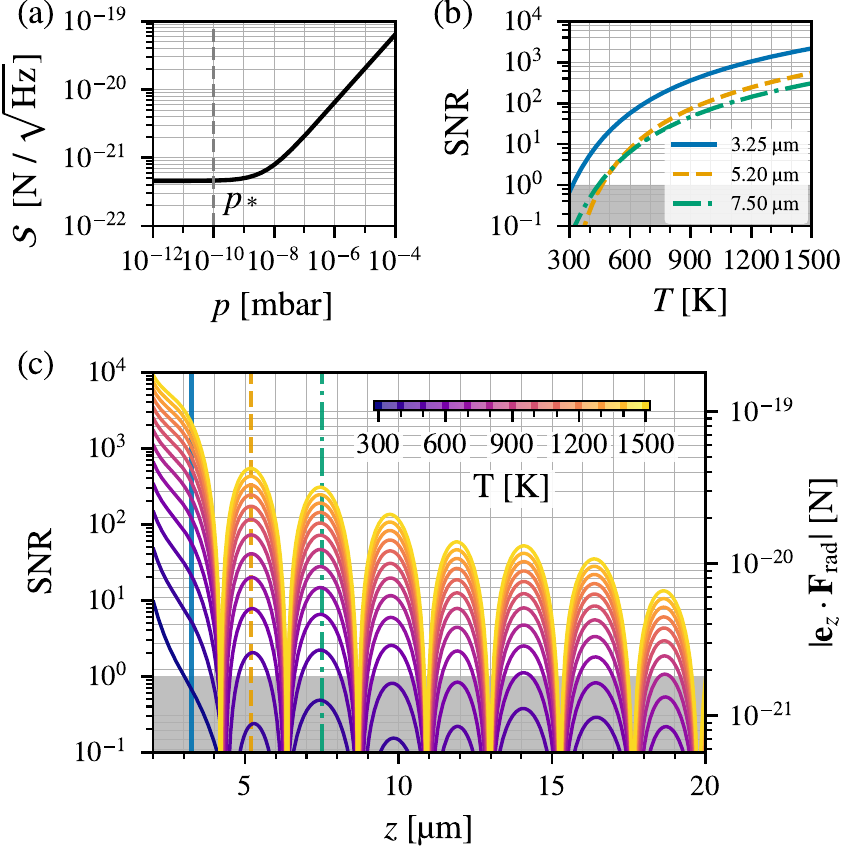}
    \caption{(a) The force sensitivity as a function of the ambient gas pressure. The vertical dashed line identifies $p_*$, the pressure below which the sensitivity is photon shot-noise limited. We use $p_*$ for plotting the other panels. (b) The signal-to-noise ratio (SNR) as a function of the nanoparticle internal temperature for three selected values of $z$. The gray area indicates the limit of measurability ($\text{SNR}<1$). (c) The force (right y-axis) and SNR (left y-axis) as a function of $z$ for $T=\SI{300}{K},\SI{400}{K},\ldots ,\SI{1500}{K}$. The grid lines follow the force axis. The vertical lines indicate the selected values of $z$ for which the full $T$-dependence is displayed in panel (b).}
    \label{fig3}
\end{figure}

In order to generate a time-oscillating $T$-dependent force that can be dynamically sensed~\cite{gonzalez-ballestero_Science_374:6564_2021}, the reflection coefficient of the surface around the wavelength $\lambda_\mathrm{peak}$ is modulated as $R(t,\lambda)=1 - \eta[1-\cos{(\Omega_\mathrm{d} t)}]/2$ (\subfigref{1}{b}). By assuming that $\partial_\lambda R(t,\lambda) = 0$ across the \SI{0.4}{\micro m} FWHM of the peak in $\Im{\alpha(\lambda)}$ at $\lambda_\mathrm{peak}$, $\vb F_\mathrm{rad}$ splits into an average force and a time-dependent force $\eta |\vb F_\mathrm{rad}(T,z)|\cos(\Omega_\mathrm{d} t)/2$ \cite{sm}. 
The result is the appearance of a peak in the nanoparticle's motional PSD at the frequency $\Omega_\mathrm{d}$ whose maximum is proportional to $|\vb F_\mathrm{rad}(T,z)|^2$.
 The $\Omega_\mathrm{d}$ frequency component of the reflection coefficient at the optical wavelength $R(t,\lambda_0)$, which we define as $\eta_0$, must simultaneously be kept small to avoid a similar and competing contribution to the motional PSD from the optical mirror image interaction force. More precisely, we require  $\eta_0/\eta < |\vb F_\mathrm{rad}|/|\vb F_\mathrm{cs}|$, where $\vb F_\mathrm{cs}$ is the force the nanoparticle would experience due to its interaction with its optical mirror image in a perfectly reflecting surface. As we show in \cite{sm}, the ratio $|\vb F_\mathrm{rad}|/|\vb F_\mathrm{cs}| \sim 10^{-4}$, and we therefore require $\eta_0/\eta < 10^{-4}$. An experimentally feasible method to achieve this is to use metasurface optical filters \cite{wang_Mater.Res.Express_5:4_2018, wang_IEEEPhotonicsJ._12:4_2020, huang_Rep.Prog.Phys._83:12_2020, kang_Adv.Opt.Mater._7:14_2019, ju_NatureNanotech_6:10_2011, komar_Appl.Phys.Lett._110:7_2017, shi_Nanomaterials_10:12_2020}, see \cite{sm} for further details.
When this requirement is fulfilled, the ratio of the $\Omega_\mathrm{d}$ peak to the thermally driven motional PSD defines the signal-to-noise ratio (SNR), which for $p \leq p_*$ becomes~\cite{sm}
\begin{equation}
    \mathrm{SNR}(\Omega_\mathrm{d}) = \frac{15 c \epsilon_0^2}{256 \pi ^4 \hbar} \frac{\lambda_0^7 Q \eta^2 |\vb F_\mathrm{rad}(T,z)|^2}{\Omega_\mathrm{d} |\alpha(\lambda_0)|^2 N^2 P_\mathrm{tw}}.
\end{equation}
Here $Q$ is the quality factor of the reflection coefficient modulation and $P_\mathrm{tw}$ is the optical tweezer power. We plot the SNR as a function of $T$ in \subfigref{3}{b} and as a function of $z$ in \subfigref{3}{c}. \subfigref{3}{b} shows that the force is measurable ($\SNR>1$) in the entire temperature range 300--\SI{1500}{K} for $z=\SI{3.25}{\micro m}$, and at least in the range 400--\SI{1500}{K} for $z=\SI{7.50}{\micro m}$. In \subfigref{3}{c}, we observe that with the chosen parameters, the force remains measurable for temperatures above \SI{800}{K} even at separations of \SI{19}{\micro m} and above. Additionally, we see in \subfigref{3}{b} that the SNR depends strongly on the internal temperature $T$, changing by several orders of magnitude (depending on $z$) as the temperature is increased by a factor of 5.

These results show that measurement of the driven $\vb F_\mathrm{rad}$ provides a way to perform thermometry of the nanoparticle internal temperature in ultrahigh vacuum~\cite{millen_NatureNanotech_9:6_2014, hoang_NatCommun_7:1_2016, delord_Appl.Phys.Lett._111:1_2017, hebestreit_Phys.Rev.A_97:4_2018, riviere_AVSQuantumSci._4:3_2022}. Due to absence of internal cooling by residual gas at $p_*$, the main heat dissipation channel of the nanoparticle will be radiative cooling. This opens up the prospect of investigating the radiative thermalization of the nanoparticle, similar to what was done for a silica nanofiber in~\cite{wuttke_Phys.Rev.Lett._111:2_2013}, but here for an isolated (i.e. unclamped) nano-sized object. 
It was recently argued that the radiative thermalization of a highly isolated nanoparticle might differ from the predictions of fluctuational electrodynamics due to failure of the local equilibrium assumption during thermalization \cite{rubiolopez_Phys.Rev.B_98:15_2018}.
Qualitatively, the internal temperature of a subwavelength silica nanoparticle with polarizability as in \subfigref{1}{c} and which is being heated by laser absorption and cooled radiatively is predicted by fluctuational electrodynamics to obey \cite{sm}
\begin{equation}
    T(t) = \left\{\begin{matrix} T_\infty\tanh{[t/\tau + \artanh(T_0/T_\infty)]} & (T_\infty > T_0) \\
    T_\infty\coth{[t/\tau + \artanh(T_\infty/T_0)]} & (T_\infty < T_0) \end{matrix}\right..
    \label{temp_dynamics}
\end{equation}
Here, $T_0$ and $T_\infty$ are the temperatures at $t=0$ and $t=\infty$ respectively, and $\tau$ is a time constant which depends only on the properties of silica glass and the optical tweezer parameters but is independent on the subwavelength particle size.
With the polarizability of \subfigref{1}{c} and parameters as in Table~\ref{tab:parameters}, we find $T_\infty \approx \SI{500}{K}$ and $\tau \approx \SI{0.2}{s}$. In experiments, $T_0$ can be independently controlled by using a dedicated heating laser~\cite{hebestreit_Phys.Rev.A_97:4_2018}.
By combining our proposed thermometry scheme with such a setup, departures from the radiative cooling described by \eqnref{temp_dynamics}, as predicted in~\cite{rubiolopez_Phys.Rev.B_98:15_2018}, could be experimentally tested. 

To conclude, we have proposed an experiment to measure an out-of-equilibrium Casimir force using an optically levitated nanoparticle in ultrahigh vacuum. We believe this is interesting {\em per se} given the challenge to measure these forces~\cite{sushkov_NaturePhys_7:3_2011, bimonte_Phys.Rev.A_92:3_2015, bimonte_Annu.Rev.Condens.MatterPhys._8:1_2017}. In addition, we have discussed how dynamic sensing of this force can be used to measure the internal temperature of the nanoparticle, both in the steady state as well as in a dynamical setting with radiative cooling taking place. This complements previous methods for measuring the internal temperatures of levitated nanoparticles, requiring either higher pressures \cite{millen_NatureNanotech_9:6_2014, hebestreit_Phys.Rev.A_97:4_2018} or nanoparticles with embedded quantum emitters \cite{hoang_NatCommun_7:1_2016, rahman_SciRep_6:1_2016, riviere_AVSQuantumSci._4:3_2022}. Despite operating in the classical regime, we view the proposed experiment as highly relevant for current efforts to prepare large quantum superposition states of a levitated nanoparticle~\cite{romero-isart_Phys.Rev.Lett._107:2_2011,romero-isart_Phys.Rev.A_84:5_2011, bateman_Nat.Commun._5:1_2014, yin_Phys.Rev.A_88:3_2013, wan_Phys.Rev.Lett._117:14_2016, romero-isart_NewJ.Phys._19:12_2017, pino_QuantumSci.Technol._3:2_2018, stickler_NewJ.Phys._20:12_2018, weiss_Phys.Rev.Lett._127:2_2021, neumeier_FastQuantum_2022}. The design of these protocols is constrained by decoherence due to thermal radiation. One can show that this decoherence rate is proportional to $\int_0^\infty \dd\lambda\, \lambda^{-7} n(T,\lambda) \Im{\alpha(\lambda)}$~\cite{hornberger_Phys.Rev.A_70:5_2004, schlosshauer_DecoherenceQuantumtoclassical_2007}, and hence critically depends on both the nanoparticle internal temperature and polarizability at infrared frequencies. Our proposed experiment gives information on both properties; the latter since the decoherence rate shows a similar dependence on $\alpha(\lambda)$ as \eqnref{Frad}. 
In addition, the observation that the thermal radiation of the nanoparticle is dominated by a narrow wavelength range would enable new strategies for the management of center-of-mass decoherence. We are currently investigating the effect on the decoherence rate of suppressing the radiation at the peak thermal wavelength. We hope this work will trigger the realization of classical experiments with levitated nanoparticles that provide key information about the physics related to sources of decoherence, which we consider pivotal in enabling the ambitious goal of preparing a nanoparticle in a large quantum superposition state.

We thank Romain Quidant for suggesting to us the idea of modulating the surface reflection coefficient. This research has been supported by the European Research Council (ERC) under the grant Agreement No. [951234] (Q-Xtreme ERC-2020-SyG). 

\nocite{novotny_PrinciplesNanoOptics_2006, wylie_Phys.Rev.A_30:3_1984, scipy1.0, piessens_QuadpackSubroutine_2012, jadidi_NanoLett._16:4_2016, gonzalez-ballestero_Phys.Rev.A_100:1_2019, messina_Phys.Rev.B_88:10_2013}

\bibliography{references.bib}

%
%
%
%
\clearpage
\newpage
\onecolumngrid
\begin{center} \textbf{\Large Supplemental material} \end{center}
\newcommand{\beginsupplement}{
    \setcounter{section}{0}
	\setcounter{table}{0}
	\setcounter{figure}{0}
	\setcounter{equation}{0}
	\setcounter{page}{1}
	\renewcommand{\thetable}{S\arabic{table}}
	\renewcommand{\thefigure}{S\arabic{figure}}
	\renewcommand{\theequation}{S\arabic{equation}}
	\renewcommand\thesection{S\arabic{section}}
	}
	
\beginsupplement

%
%
\section{\label{s:force_derivation} Derivation of the forces acting on the nanoparticle}

In this section we give further details of the derivation of the electromagnetic forces acting on the nanoparticle that are discussed in the main text.
\newcommand{\gpv}{\vb r} 

We consider a point dipole at position $\dpv$ with electric dipole moment given by $\vb d(t)$. The dipole is placed in vacuum and possibly in the vicinity of dielectric material (e.g.\ a surface). The total electromagnetic field in the space is given by $\vb E(\gpv,t)$. The electric dipole moment has a contribution $\vb d_\mathrm{th}(t)$ from intrinsic thermal fluctuations at temperature $T$ (e.g.\ due to the internal temperature of a nanoparticle) and a contribution $\vb d_\mathrm{ind}(t)$ due to the dipole moment induced by the external electric field. The electromagnetic field $\vb E(\gpv,t)$ has a contribution $\vb E_\mathrm{th}(\gpv, t)$ from environmental thermal fluctuations at temperature $\Tenv$ (i.e.\ due to the temperature of the vacuum and the surrounding dielectric material), a contribution $\vb E_\mathrm{tw}(\gpv,t)$  from the laser beam used for trapping, and a contribution $\vb E_\mathrm{ind}(\gpv, t)$ from the field emitted by the point dipole. In total, we have 
\begin{align}
    \vb d(t) & = \vb d_\mathrm{th}(t) + \vb d_\mathrm{ind}(t) \label{seq:total_fluc_dipole}, \\
    \vb E(\gpv, t) & = \vb E_\mathrm{th}(\gpv, t) + \vb E_\mathrm{tw}(\gpv, t) + \vb E_\mathrm{ind}(\gpv, t). \label{seq:total_fluc_field}
\end{align}
In order to describe how the induced electric dipole moment depends on the electromagnetic field and vice versa, it is convenient to transform to frequency domain, defined by
\begin{align}
    \vb d(\omega) & = \int_{-\infty}^\infty \dd t  \, \vb d(t) \ee^{\ii \omega t}, \label{seq:freq_dipole_def} \\
    \vb E(\vb r, \omega) & = \int_{-\infty}^\infty \dd t  \, \vb E(\vb r, t) \ee^{\ii \omega t}\, . \label{seq:freq_field_def}
\end{align}
The electromagnetic field emitted by the point dipole can be expressed as
\begin{equation}\label{seq:induced_field}
    \vb E_\mathrm{ind}(\gpv, \omega) = \tensor{G}(\gpv, \dpv, \omega)\vb d(\omega) = \tensor{G}_0(\gpv, \dpv, \omega)\vb d(\omega) + \tensor G_1(\gpv, \dpv, \omega)\vb d(\omega).
\end{equation}
Here $\tensor{G}(\gpv, \dpv, \omega)$ is Green's tensor for the electromagnetic field that describes the field at position $\gpv$ emitted by the point dipole at position $\dpv$ in the presence of the surrounding material. In the second equality we used $\tensor{G}(\gpv, \dpv, \omega) = \tensor{G}_0(\gpv, \dpv, \omega) + \tensor{G}_1(\gpv, \dpv, \omega)$, where $\tensor{G}_0(\gpv, \dpv, \omega)$ is the Green's tensor in vacuum, namely in absence of any material. The Green's tensor $\tensor{G}_1(\gpv, \dpv, \omega) $ is often referred to as the scattering part of the total Green's tensor.

The induced electric dipole moment can be expressed as
\begin{equation}\label{seq:induced_dipole}
    \begin{aligned}[b]
        \vb d_\mathrm{ind}(\omega) & = \alpha(\omega)\vb E_\mathrm{ex}(\dpv,\omega) \\
        & = \alpha(\omega) \left[ \vb E(\dpv, \omega) - \tensor{G}_0(\dpv, \dpv,\omega)\vb d(\omega) \right] \\ 
        & = \alpha(\omega) \left[ \vb E_\mathrm{th}(\dpv, \omega) + \vb E_\mathrm{tw}(\dpv, \omega) + \tensor{G}_1(\dpv, \dpv,\omega)\vb d(\omega) \right].
    \end{aligned}
\end{equation}
Here $\alpha(\omega)$ is the scalar polarizability of the electric point dipole. The electromagnetic field inducing the dipole is given by the so called exciting field defined as $\vb E_\mathrm{ex}(\gpv,\omega) \equiv \vb E(\gpv, \omega) - \tensor{G}_0(\gpv, \dpv,\omega)\vb d(\omega)$. In this way, the dipole can be induced either by the external thermal field, the laser beam, or by the field emitted by the dipole that is reflected by the nearby material.

The instantaneous electromagnetic force acting on the point dipole can be evaluated as~\cite{novotny_PrinciplesNanoOptics_2006} 
\begin{equation}  
        \vb F(t) =  \sum_{j\in\{x,y,z\}} d_j(t) \left.\nabla_{\gpv} E_{\mathrm{ex},j}(\gpv, t)\right|_{\gpv=\dpv} + \dfdx{}{t}\left[\vb d(t) \cdot \vb B(\dpv, t)\right]  = \nabla_{\gpv} \left[\vb d(t) \cdot \vb E_{\mathrm{ex}}(\gpv, t)\right]_{\gpv=\dpv} + \dfdx{}{t}\left[\vb d(t) \cdot \vb B(\dpv, t)\right].
\end{equation}
 This is a fluctuating force due to the thermal fluctuations of the dipole $\vb d_\mathrm{th}(t)$ and the electric field $\vb E_\mathrm{th}(\gpv, t)$. We are interested in the time-averaged value of the force $\vb F = \expval{\vb F(t)}$. The thermally fluctuating quantities are assumed to be both stationary and ergodic, hence the time average is equivalent to an ensemble average. Stationarity implies that the total time derivative of the second term in $\vb F(t)$ is zero under time average. Thus, one has
\begin{equation}   \label{seq:definition_force}
        \vb F = \expval{\nabla_{\gpv} \left [ \vb d(t) \cdot \vb E_{\mathrm{ex}}(\gpv, t)\right ]_{\gpv=\dpv}}  =  \frac{1}{4 \pi^2} \int_{-\infty}^{\infty} \int_{-\infty}^{\infty} \dd \omega  \dd \omega'  \left.\nabla_{\gpv} \expval{\vb d(\omega) \cdot \vb E^*_{\mathrm{ex}}(\gpv, \omega')}\right|_{\gpv=\dpv},  
\end{equation}
where the quantity $\expval{\vb d(\omega) \cdot \vb E^*_{\mathrm{ex}}(\gpv, \omega')}$ is the ensemble average over the realisations of the fluctuations in frequency space.

In order to evaluate the frequency correlators in the integrand of \eqnref{seq:definition_force}, we insert \eqnref{seq:induced_field} and \eqnref{seq:induced_dipole} into the frequency-domain versions of \eqnref{seq:total_fluc_dipole} and \eqnref{seq:total_fluc_field} and solve for $\vb d(\omega)$ and $\vb E(\gpv, \omega)$. The result is
\begin{align}
        \label{seq:total_dipole_from_fluc}
        \vb d(\omega) & = \dresstensor(\dpv,\omega)\left\{ \vb d_\mathrm{th}(\omega) + \alpha(\omega) [\vb E_\mathrm{th}(\dpv, \omega) + \vb E_\mathrm{tw}(\dpv, \omega)]\right\} ,\\
        \label{seq:ex_field_from_fluc}
        \vb E_\mathrm{ex}(\gpv,\omega) & = \vb E_\mathrm{th}(\gpv,\omega) + \vb E_\mathrm{tw}(\gpv, \omega) + \tensor G_1(\gpv,\dpv,\omega)\dresstensor(\dpv, \omega)\left\{\vb d_\mathrm{th}(\omega) + \alpha(\omega)[\vb E_\mathrm{th}(\dpv,\omega) + \vb E_\mathrm{tw}(\dpv,\omega)]\right\},
\end{align}
where
\begin{equation}
    \dresstensor (\dpv,\omega) \equiv \left [ \ut - \alpha (\omega) \tensor G_1(\dpv, \dpv, \omega)\right ]^{-1}.
    \label{seq:dresstensor_def}
\end{equation}
The tensor $\tensor T$ accounts for the ability of the dipole to scatter the radiation that it previously has emitted and that returns to the dipole after having been reflected by the environment (e.g.\ surface).
 
 To understand the action of $\tensor T$ in more detail, we can introduce the ``bare'' dipole moment $\vb d^{(0)}$ as
 \begin{equation}
    \vb d^{(0)}(\omega) = \vb d_\mathrm{th}(\omega) + \alpha(\omega)[\vb E_\mathrm{th}(\dpv, \omega) + \vb E_\mathrm{tw}(\dpv, \omega)].
 \end{equation}
 Next, recall that $\tensor G \vb d^{(0)}$ is the field radiated by the bare dipole, and in particular, that $\tensor G_1 \vb d^{(0)}$ is the radiation from the bare dipole that is subsequently reflected by the environment. The dipole moment that this field induces when it returns to the dipole is $\alpha \tensor G_1 \vb d^{(0)}$. Therefore, we introduce the $n\text{:th}$ induced dipole moment as
 \begin{equation}
    \vb d^{(n)}(\omega) = \alpha(\omega)\tensor G_1(\dpv,\dpv,\omega)\vb d^{(n-1)}(\omega).
    \label{seq:nth_induced_dipole}
 \end{equation}
 In words, the dipole $\vb d^{(n)}$ is the dipole moment that gets induced when radiation from the dipole $\vb d^{(n-1)}$ gets reflected by the environment and returns to the location of the point dipole. Then we see that \eqnref{seq:total_dipole_from_fluc} can be formally written as
 \begin{equation}
    \vb d(\omega) = \dresstensor(\dpv,\omega)\vb d^{(0)}(\omega) = \vb d^{(0)}(\omega) + \sum_{n=1}^\infty \vb d^{(n)}(\omega).
 \end{equation} 
 In words, the total dipole moment is the bare dipole moment plus all the dipole moments that get induced in sequence as multiple reflections between the dipole and the environment happen. 
 We then see that \eqnref{seq:ex_field_from_fluc} states that the total field has independent contributions from the thermal field, laser field, and the field radiated by the dipole including all the multiple scattering events of the dipole.
 
 For sufficiently small polarizability, namely for $||\alpha \tensor G_1|| \ll 1$, one can neglect the contributions of the induced dipole moments to the total dipole moment, i.e.~the dipole's rescattering of its own radiation can be neglected. In this case, expressions of the forces can be approximated by using $\dresstensor \approx \ut$. 
 Under this approximation, the field emitted by the bare dipole only gets reflected once -- by the environment. Roughly speaking, the dipole feels the force from the radiation reflected by the environment, but it does not emit additional radiation due to the interaction. We will use this approximation later.
 
The frequency correlators in \eqref{seq:definition_force} can now be evaluated in terms of the frequency correlators of the fluctuating and driving fields using \eqnref{seq:total_dipole_from_fluc} and \eqnref{seq:ex_field_from_fluc}. We assume that the cross-correlations of $\vb E_\mathrm{th}$, $\vb E_\mathrm{tw}$, and $\vb d_\mathrm{th}$ are zero, so that any expectation value with a combination of these quantities vanishes. Therefore, the only terms that are nonzero are those containing any of the following three correlators:
\begin{gather}
    \begin{aligned}[b]
        \expval{d_{\mathrm{th},j}(\omega) d_{\mathrm{th},k}^*(\omega')} = 2\pi\hbar\delta(\omega-\omega') \coth{\left(\frac{\hbar\omega}{2\boltzmann T}\right)} \delta_{jk} \Im{\alpha(\omega)},
        \label{seq:fdrel_dipole}
    \end{aligned}
    \\
    \begin{aligned}[b]
        \expval{E_{\mathrm{th},j}(\dpv,\omega) E^*_{\mathrm{th},k}(\gpv,\omega')} = 2\pi\hbar\delta(\omega-\omega')\coth{\left(\frac{\hbar\omega}{2\boltzmann \Tenv}\right)}\Im{G_{jk}(\dpv, \gpv, \omega)},
        \label{seq:fdrel_field}
    \end{aligned}
    \\
    \expval{E_{\mathrm{tw},j}(\dpv, \omega) E_{\mathrm{tw},k}^*(\gpv, \omega')} = \delta(\omega-\omega') \lim_{\Delta\omega\to0}\int_{\omega - \Delta\omega}^{\omega + \Delta\omega} \dd\omega''\,E_{\mathrm{tw},j}(\dpv, \omega) E_{\mathrm{tw},k}^*(\gpv, \omega'').
    \label{seq:cfield_psd}
\end{gather}
The thermal correlators \eqnref{seq:fdrel_dipole} and \eqnref{seq:fdrel_field} are obtained using the fluctuation-dissipation theorem at the temperature of the point dipole and the electromagnetic field environment respectively. In the nonequilibrium situation $T \neq \Tenv$ (i.e.\ when the point dipole has a different temperature than its environment), this assumes local equilibrium. The relation \eqnref{seq:cfield_psd} holds for any field which is stationary after a time-average is performed. In the formulas that follow, we make the integral in \eqnref{seq:cfield_psd} implicit by defining:
\begin{equation}
    \lim_{\Delta\omega\to0}\int_{\omega - \Delta\omega}^{\omega + \Delta\omega} \dd\omega''\,E_{\mathrm{tw},j}(\dpv, \omega) E_{\mathrm{tw},k}^*(\gpv, \omega'') \equiv E_{\mathrm{tw},j}(\dpv, \omega) \bar E_{\mathrm{tw},k}^*(\gpv, \omega).
\end{equation}

Putting everything together, we identify five contributions to the mean force:
\begin{equation}
    \vb F = \vb F_\mathrm{trap} + \vb F_\mathrm{cs} + \vb F_\mathrm{zpf} + \vb F_\mathrm{env} + \vb F_\mathrm{rad}.
    \label{seq:total_force}
\end{equation}
For each term on the right-hand side, we now give the general expression of the term, the expression within the $\dresstensor\to \ut$ approximation, and a physical explanation of the term's appearance.

(1) The first term $\vb F_\mathrm{trap}$ is given by
\begin{equation}
    \begin{aligned}[b]
        \vb F_\mathrm{trap} & = \begin{aligned}[t]
            \int_{0}^\infty\frac{\dd\omega}{2\pi^2} \nabla_{\gpv} 
            \Bigl[ & \Re{\alpha(\omega)}\Re{\bar{\vb E}_\mathrm{tw}^*(\vb r,\omega) \cdot \dresstensor(\vb r_0, \omega) \vb E_\mathrm{tw}(\dpv, \omega)}_{\gpv=\dpv} \\
            & + \Im{\alpha(\omega)}\Im{\bar{\vb E}_\mathrm{tw}^*(\vb r,\omega) \cdot \dresstensor(\vb r_0, \omega) \vb E_\mathrm{tw}(\dpv, \omega)}_{\gpv=\dpv}\Bigr] 
        \end{aligned} \\
        & \approx \int_{0}^\infty\frac{\dd\omega}{2\pi^2} \left[\frac{1}{2} \Re{\alpha(\omega)} \nabla_{\dpv} \left[\bar{\vb E}^*_\mathrm{tw}(\dpv,\omega) \cdot \vb E_\mathrm{tw}(\dpv,\omega)\right] + \Im{\alpha(\omega)} \nabla_\gpv \Im{\bar{\vb E}^*_\mathrm{tw}(\dpv,\omega) \cdot \vb E_\mathrm{tw}(\gpv,\omega)}_{\gpv=\dpv}\right] \quad\quad ({\dresstensor \to \ut}).
    \end{aligned}
    \label{seq:trap_force}
\end{equation}
This term describes the optical gradient and scattering force exerted by the laser beam. In our proposal, this term will be used to trap the particle (e.g. optical tweezer). We remark that the field $\vb E_\mathrm{tw}$ describes the laser beam in the presence of the surrounding material but in the absence of the dipole.

(2) The second term $\vb F_\mathrm{cs}$ is given by
\begin{equation}
    \begin{aligned}[b]
        \vb F_\mathrm{cs} & = \int_{0}^\infty\frac{\dd\omega}{2\pi^2}\,|\alpha(\omega)|^2\nabla_\gpv\Re{\bar{\vb E}_\mathrm{tw}^*(\dpv,\omega) \cdot \dresstensor^\dagger(\dpv,\omega)\tensor G_1^\dagger(\gpv,\dpv,\omega) \dresstensor(\dpv,\omega) \vb E_\mathrm{tw}(\dpv,\omega)}_{\gpv=\dpv} \\
        & \approx \int_{0}^\infty\frac{\dd\omega}{2\pi^2}\, |\alpha(\omega)|^2 \nabla_{\gpv}\Re{\bar{\vb E}_\mathrm{tw}^*(\dpv,\omega) \cdot \tensor G_1^\dagger(\gpv, \dpv, \omega) \vb E_\mathrm{tw}(\dpv,\omega)}_{\gpv=\dpv} \quad\quad (\dresstensor \to \ut).
        \label{seq:cs_force}
    \end{aligned}
\end{equation}
This force arises due to the nanoparticle scattering light from the trapping laser beam, which is subsequently reflected from the environment, and returns to interact with the nanoparticle's dipole moment. This is very similar to the force observed in~\cite{rieser_Science_377:6609_2022}, but here the interaction is between the nanoparticle and its optical mirror image rather than between two distinct nanoparticles. Note that the force depends on the orientation of the electric field at the position of the dipole, that is, on the polarization of the laser beam. 

(3) The third term $\vb F_\mathrm{zpf}$ is given by
\begin{equation}
    \begin{aligned}[b]
        \vb F_\mathrm{zpf} & = \begin{aligned}[t]
            \int_0^\infty \frac{\hbar \dd\omega}{\pi} \nabla_\gpv \mathrm{Tr}\biggl\{ & \Re{\alpha(\omega)\dresstensor(\dpv,\omega)} \Im{\tensor G(\dpv, \gpv,\omega)} \\    
            & + \Re{\dresstensor^\dagger(\dpv,\omega) \tensor G_1(\gpv, \dpv,\omega)\dresstensor(\dpv,\omega)}\left(\Im{\alpha(\omega)} + |\alpha(\omega)|^2\Im{\tensor G(\dpv, \dpv,\omega)}\right)\biggr\}_{\gpv=\dpv}
        \end{aligned}\\
        & \approx \int_0^\infty \frac{\hbar \dd\omega}{2\pi} \left\{\Im{\alpha(\omega) \nabla_{\dpv} \tr\left\lbrace\tensor G_1(\vb r_0, \vb r_0, \omega)\right\rbrace} + 2|\alpha(\omega)|^2\nabla_\gpv\operatorname{Tr}\left[\Re{\tensor G_1(\gpv, \dpv,\omega)}\Im{\tensor G(\dpv, \dpv,\omega)}\right]_{\gpv=\dpv}\right\}.
    \end{aligned}
    \label{seq:zpf_force}
\end{equation}
This force arises due to the quantum zero-point fluctuations of the dipole moment of the nanoparticle correlating with the zero-point fluctuations of the environmental electromagnetic field.  
In the $\dresstensor \to \ut$ approximation, the first and second terms combine into the $\Im{\alpha \tr\left\{\tensor G_1\right\}}$ term. It is typically this contribution to the force that is referred to by the name zero-temperature Casimir--Polder force~\cite{wylie_Phys.Rev.A_30:3_1984}.

(4) The fourth term $\vb F_\mathrm{env}$ contains all the dependence on the environmental temperature $\Tenv$ and is given by
\begin{equation}
    \begin{aligned}[b]
        \vb F_\mathrm{env} & =
        \begin{aligned}[t]
            \int_0^\infty \frac{2\hbar \dd\omega}{\pi} n(\Tenv,\omega)\nabla_\gpv \mathrm{Tr}\biggl\{ & \Re{\alpha(\omega)\dresstensor(\dpv,\omega)} \Im{\tensor G(\dpv, \gpv,\omega)} \\
            & + |\alpha(\omega)|^2\Re{\dresstensor^\dagger(\dpv,\omega) \tensor G_1(\gpv, \dpv,\omega)\dresstensor(\dpv,\omega)}\Im{\tensor G(\dpv, \dpv,\omega)}\biggr\}_{\gpv=\dpv}
        \end{aligned} \\
        & \begin{aligned}[b]
            \approx \int_0^\infty \frac{\hbar \dd\omega}{\pi} n(T_\mathrm{env}, \omega) \Bigl\{
                & \Re{\alpha(\omega)}\Im{ \nabla_{\dpv} \tr\left\{\tensor G(\vb r_0, \vb r_0, \omega)\right\}} \\
                & + 2|\alpha(\omega)|^2\nabla_\gpv \tr\Bigl[\Re{\tensor G_1(\gpv, \dpv,\omega)}\Im{\tensor G(\dpv, \dpv,\omega)}\Bigr]_{\gpv=\dpv}\Bigr\} \quad\quad (\dresstensor\to\ut).
            \end{aligned}
    \end{aligned}
    \label{seq:env_force}
\end{equation}
This force arises due to environmental thermal radiation being scattered by the dipole. The first term in the force describes a gradient-type force on the total dipole due to the gradient of the thermal field intensity, and the second term is the thermal version of $\vb F_\mathrm{cs}$, where the dipole scatters light from the thermal field which is then reflected and returns to interact with the dipole.

(5) Finally, the fifth term $\vb F_\mathrm{rad}$ contains all dependence on the particle temperature $T$ and is given by
\begin{equation}
    \begin{aligned}[b]
        \vb F_\mathrm{rad} & = \int_0^\infty\frac{2\hbar\dd\omega}{\pi}\, n(T,\omega) \Im{\alpha(\omega)}\nabla_{\gpv}\Re{\tr\left[\dresstensor^\dagger(\dpv,\omega) \tensor G_1^\dagger(\gpv,\dpv,\omega) \dresstensor(\dpv,\omega)\right]}_{\gpv=\dpv} \\
        & \approx \int_0^\infty\frac{\hbar\dd\omega}{\pi}\, n(T,\omega)\Im{\alpha(\omega)}\nabla_{\dpv}\Re{\tr\left [\tensor G_1^\dagger(\dpv,\dpv,\omega)\right]} \quad\quad (\dresstensor\to\ut).
    \end{aligned}
    \label{seq:bb_force}
\end{equation}
where $n(T,\omega) = [\exp{(\hbar\omega/\boltzmann T)} - 1]^{-1}$ and in the second line we used $\nabla_\gpv \tr[\tensor G(\vb r,\vb r_0)] =  \nabla_{\dpv} \tr[ \tensor G(\vb r_0,\vb r_0)]/2$.
This force arises due to thermal radiation emitted by the dipole which is reflected by the environment and returns to interact with the dipole. 
In order for this force to be non-vanishing, the thermal radiation that returns to the dipole must retain some coherence with the dipole. This is possible for \emph{colored} thermal emitters, meaning that $\Im{\alpha(\omega)}$ is peaked in the infrared part of the spectrum. We emphasize that the gradient and scattering force that the dipole would experience due to its reflected field are not included in the $\dresstensor \to \ut$ expression, but are part of negligible higher-order terms. To see this, it is useful to use the $n$:th induced thermal dipole moment and field akin to \eqnref{seq:nth_induced_dipole}: $\vb d_\mathrm{th}^{(n)} = \alpha \vb E_\mathrm{th}^{(n-1)}$ and $\vb E^{(n)}_\mathrm{th} = \tensor G_1 \vb d_\mathrm{th}^{(n)}$. Then,
\begin{equation}
        \vb F_\mathrm{rad} \propto \expval{\vb d_\mathrm{th}^* \cdot \tensor G_1 \vb d_\mathrm{th}} = \expval{\left(\dresstensor \vb d^{(0)}_\mathrm{th}\right)^* \cdot \tensor G_1 \dresstensor \vb d^{(0)}_\mathrm{th}}
        = \expval{{\vb d^{(0)}_\mathrm{th}}^* \cdot \vb E^{(0)}_\mathrm{th}} + \expval{{\vb d^{(0)}_\mathrm{th}}^*\cdot \vb E^{(1)}_\mathrm{th}} + \alpha\expval{{\vb E^{(0)}_\mathrm{th}}^* \cdot \vb E^{(0)}_\mathrm{th}} + \dots
\end{equation}
where in the last expression we included terms up to first order in $\vb d^{(1)}_\mathrm{th}$. The term ${\vb d^{(0)}_\mathrm{th}}^*\cdot \vb E^{(0)}_\mathrm{th}$ is the coherent interaction included in the $\dresstensor \to \ut$ expression, the term ${\vb d^{(0)}_\mathrm{th}}^*\cdot \vb E^{(1)}_\mathrm{th}$ is the next-order coherent interaction term, while the term $\alpha{\vb E^{(0)}_\mathrm{th}}^*\cdot \vb E^{(0)}_\mathrm{th}$ is the lowest-order gradient and scattering force on the nanoparticle due to the thermal emission. As we will see later, in Fig.~\ref{sfig:small_limit_check}, the last two terms are negligible compared to the first term.

%
%
\section{Evaluation of forces for a nanoparticle in a plane geometry}
\begin{figure}
    \centering
    \includegraphics{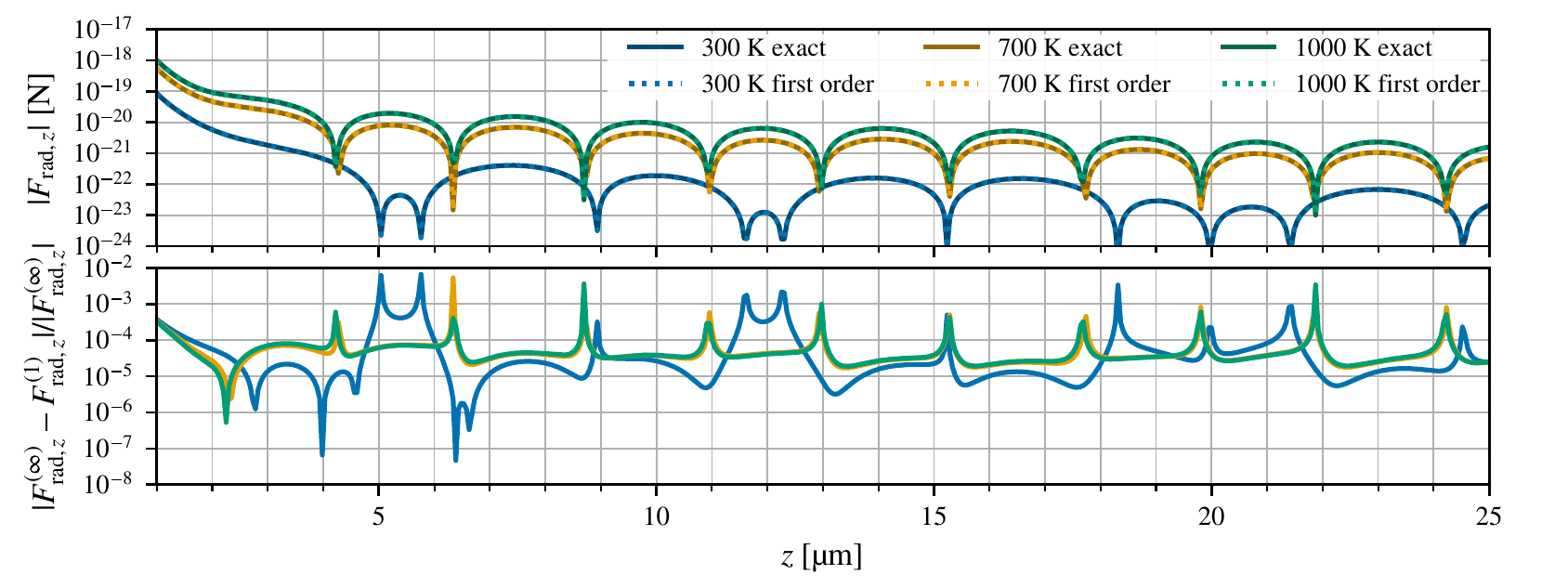}
    \caption{Comparison of $\vb F_\mathrm{rad}$ calculated using the $\dresstensor \neq \ut$ expression ($\vb F_\mathrm{rad}^{(\infty)}$) and the $\dresstensor = \ut$ expansion  ($\vb F_\mathrm{rad}^{(1)}$) of \eqnref{seq:bb_force}. Top panel: Darker solid lines are the exact result and lighter dotted lines are the $\dresstensor = \ut$ results. Bottom panel: the difference of these results as a fraction of $\vb F_\mathrm{rad}^{(\infty)}$ is displayed with colors matching temperatures in the top panel.}
    \label{sfig:small_limit_check}
\end{figure}

We explain how we evaluate the forces in the particular scenario of an optically trapped nanoparticle in front of a surface. The scattering Green's function above a planar interface is given by~\cite{buhmann_DispersionForcesI_2012}
\begin{equation}
    \tensor G_1(\vb r, \vb r',\omega) = \frac{\ii}{8\pi^2 \epsilon_0} \int_0^\infty \dd k_\parallel \frac{k_\parallel}{k_z} \int_0^{2\pi}\dd\theta\,\exp{\left\{\ii k_\parallel (\cos{\theta} \uv x + \sin{\theta}\uv y) \cdot (\vb r - \vb r') + \ii k_z(z+z')\right\}}\tensor M(k_\parallel,\theta,\omega),
    \label{seq:plane_scattering_gf}
\end{equation}
with $k_z = \left[(\omega/c)^2 - k_\parallel^2\right]^{1/2}\ (\Im{k_z}>0)$, and
\begin{equation}
    \tensor M(k_\parallel,\theta,\omega) = \left(\begin{array}{ccc}
        k^2r_s \sin^2{\theta} -k_z^2 r_p\cos^2\theta  & -k^2r_s \sin\theta\cos\theta  - k_z^2 r_p\sin \theta \cos \theta   &
        -k_p k_z r_p\cos\theta  \\
        -k^2 r_s\sin\theta\cos\theta  - k_z^2 r_p\sin\theta\cos\theta  & k^2 r_s \cos^2\theta  - k_z^2 r_p\sin^2\theta  &
        -k_p k_z r_p \sin\theta  \\
        k_p k_z r_p\cos\theta  &  k_p k_z r_p\sin\theta & k_p^2 r_p \\
    \end{array}\right),
    \label{seq:plane_scattering_M}
\end{equation}        
where $r_{s,p}=r_{s,p}(k_\parallel,\omega)$ are the Fresnel reflection coefficients for $s$ and $p$ polarizations. $\tensor G_1(\vb r_0, \vb r_0, \omega)$ becomes a diagonal matrix after evaluation of the $\theta$-integral, since $\int_0^{2\pi}\tensor M(k_\parallel,\theta,\omega) \,\dd\theta \equiv \overline{\tensor M}(k_\parallel,\omega)$ is a diagonal matrix. Consequently, $\dresstensor(\dpv,\omega) = [1 - \alpha \tensor G_1(\dpv, \dpv, \omega)]^{-1}$ is also a diagonal matrix, and the trace in \eqnref{seq:bb_force} can be evaluated as the trace of $\tensor G_1$ weighted with the components of the diagonal matrices:
\begin{equation}
    \tr\left\{\tensor T^\dagger(\dpv,\omega) \tensor G_1(\gpv, \dpv, \omega)^\dagger \tensor T(\dpv,\omega)\right\} = \sum_{j\in\{x,y,z\}} G^*_{1,jj}(\gpv,\dpv,\omega) |T_{jj}(\dpv,\omega)|^2.
\end{equation}
Now,
\begin{equation}
    \begin{aligned}[b]
        \lim_{\gpv \to \dpv}\nabla_{\gpv}G_{1,jj}(\gpv, \dpv, \omega) & = \frac{\ii}{8\pi^2 \epsilon_0} \int_0^\infty \dd k_\parallel\frac{k_\parallel}{k_z}  \ee^{2 \ii k_z z_0} \int_0^{2\pi}\dd\theta \left(\begin{matrix} \ii k_\parallel \cos{\theta} \\ \ii k_\parallel \sin{\theta} \\ \ii k_z
        \end{matrix}\right) M_{jj}(k_\parallel,\theta,\omega) \\
        & = \frac{-\uv z}{8\pi^2 \epsilon_0} \int_0^\infty \dd k_\parallel\, k_\parallel \ee^{2 \ii k_z z_0} \overline{M}_{jj}(k_\parallel,\omega),
    \end{aligned}
    \label{seq:plane_diagonal_derivative_evaluation}
\end{equation}
since the $\uv x$ and $\uv y$ components in the middle equation integrate to zero. We remark that, as the right-hand-side of \eqnref{seq:plane_diagonal_derivative_evaluation} is equal to $\nabla_{\dpv} G_{1,jj}(\vb r_0,\vb r_0)/2$, this is an explicit demonstration of the general relationship $\lim_{\vb r \to \vb r_0}\nabla_{\vb r} G_{1,jj}(\vb r, \vb r_0, \omega) = \nabla_{\dpv} G_{1,jj}(\vb r_0, \vb r_0, \omega)/2$ that we used to simplify the $\dresstensor=1$ expressions in Section~\ref{s:force_derivation}. Consequently,
\begin{multline}
    \lim_{\gpv \to \dpv}\nabla_{\gpv} \tr\left\{\Re{\tensor T^\dagger(\dpv,\omega) \tensor G_1(\gpv, \dpv, \omega)^\dagger \tensor T(\dpv,\omega)}\right\} \\ = \frac{-\uv z}{8\pi^2 \epsilon_0} \sum_{j\in\{x,y,z\}} \Re{\int_0^\infty \dd k_\parallel\, k_\parallel \ee^{2\ii k_z z_0} \overline{M}_{jj}(k_\parallel,\omega)} |T_{jj}(\dpv,\omega)|^2,
    \label{seq:diagonal_trace_derivative}
\end{multline}
with
\begin{gather}
    \overline M_{xx}(k_\parallel,\omega) = \overline M_{yy}(k_\parallel,\omega) = \pi((\omega/c)^2 r_s(k_\parallel,\omega) - k_z^2(k_\parallel) r_p(k_\parallel,\omega)),\quad \overline M_{zz}(k_\parallel,\omega) = 2\pi k_\parallel^2 r_p(k_\parallel,\omega), \label{seq:barM_general} \\
    \left(T_{jj}\right)^{-1} = 1-\frac{\ii\alpha(\omega)}{8\pi^2\epsilon_0}\int_0^\infty \dd k_\parallel\,\frac{k_\parallel}{k_z}\,  \ee^{2\ii k_z z_0} \overline{M}_{jj}(k_\parallel,\omega).
    \label{seq:Tjj_general}
\end{gather}
Insertion of Eqs.~(\ref{seq:diagonal_trace_derivative}-\ref{seq:Tjj_general}) into \eqnref{seq:bb_force} would now allow numerical evaluation of the integrals. However, we proceed to the special case of a perfectly reflecting plane which has $r_s(k_\parallel,\omega)=-1,\,r_p(k_\parallel,\omega)=1$. In this case the integrals over $k_\parallel$ can be performed analytically, and we obtain, with $k=\omega/c$
\begin{gather}
    \lim_{\vb r\to\vb r_0}\nabla_\vb r G_{1,jj}(\vb r,\vb r_0,\omega) = -\frac{\uv z \ee^{2\ii k z_0}}{64\pi \epsilon_0 z_0^4}\left\{
        \begin{array}{ll}
            3 - 6 \ii k z_0 - 8 k^2 z_0^2 + 8 \ii k^3 z_0^3 \quad \ & j=x,y \\
            6 - 12 \ii k z_0 - 8 k^2 z_0^2 \quad \ & j=z
        \end{array}
    \right. \label{seq:nablaG_jj_mirror}\\
        \left(T_{jj}(\dpv,\omega)\right)^{-1} = \left\{
            \begin{array}{ll}
                1+\alpha(\omega)\ee^{2 \ii k z_0} \left(-1 + 2 \ii k z_0 + 4 k^2 z_0^2\right)/(32 \pi  \epsilon_0 z_0^3) \quad \ & j=x,y \\
                1+\ii\alpha(\omega)  \ee^{2 \ii k z_0} (\ii + 2 k z_0)/(16 \pi  \epsilon_0 z_0^3) \quad \ & j=z
            \end{array}
        \right. \label{seq:T_jj_mirror}
\end{gather}
Usage of these relations with \eqnref{seq:diagonal_trace_derivative} leaves only the $\omega$ integral to be evaluated in \eqnref{seq:bb_force}. We use these relations to investigate the accuracy of the $\dresstensor=1$ version of \eqref{seq:bb_force} to the expression including $\dresstensor$. The $\dresstensor=1$ expansions of Eqs.~(\ref{seq:cs_force}-\ref{seq:zpf_force}) are conveniently evaluated for a perfectly reflecting plane by using
\begin{equation}
    \nabla_{\dpv} \tr\left\{\tensor G_1(\vb r_0, \vb r_0, \omega)\right\} = \frac{\uv z}{{8 \pi \epsilon_0 z_0^4}}\ee^{2 \ii k z_0} \left(-3 + 6 \ii k z_0 + 6 k^2 z_0^2 - 4 \ii k^3 z_0^3\right)
    \label{seq:plane_tr_nabla_G}
\end{equation}
which can be obtained from \eqnref{seq:nablaG_jj_mirror} combined with $\lim_{\vb r \to \vb r_0}\nabla_{\vb r} G_{1,jj}(\vb r, \vb r_0, \omega) = \nabla_{\dpv} G_{1,jj}(\vb r_0, \vb r_0, \omega)/2$, or from the method of images. We compare the $\dresstensor\neq\ut$ and $\dresstensor=\ut$ results for $\vb F_\mathrm{rad}$ in Figure~\ref{sfig:small_limit_check}. We find agreement between the full and expanded results, as expected for a subwavelength particle since it fulfills $||\alpha \tensor G_1||\ll 1$. Therefore we use the $\dresstensor=\ut$ versions of all forces in the rest of this work.

We evaluate the total force on the nanoparticle $\vb F$ displayed in \figref{2} in the main text as follows: Starting from \eqref{seq:total_force}, we neglect $\vb F_\mathrm{trap}$ since its later effect on the PSD is captured by the trapping frequency $\Omega$ and equilibrium position $\vb r_0$. $\vb F_\mathrm{rad}$ and $\vb F_\mathrm{env}$ were calculated using \eqref{seq:bb_force} and \eqref{seq:env_force} with \eqref{seq:plane_tr_nabla_G} and $\alpha(\omega)$ as displayed in \subfigref{1}{c} in the main text. In \eqref{seq:env_force}, we neglect the term $\propto |\alpha(\omega)|^2$. We performed the integrals numerically using the Gauss--Kronrod 21-point rule as implemented in the \texttt{scipy.integrate.quad\_vec} method of SciPy 1.8.0~\cite{scipy1.0, piessens_QuadpackSubroutine_2012}. 
As integration bounds we used the upper and lower bounds $\lambda_\mathrm{max}=\SI{50}{\micro m}$ and $\lambda_\mathrm{min}=\SI{7}{\micro m}$ of the $\epsilon(\lambda)$ dataset~\cite{kitamura_Appl.Opt._46:33_2007}, translated into frequency space as $\omega_\mathrm{min,max} = 2\pi c/\lambda_\mathrm{max,min}$. We evaluate $\vb F_\mathrm{zpf}$ with the far-field approximation of the Casimir force 
\begin{equation}
    \vb F_\mathrm{zpf}(z \gg \nprad) = -\frac{\hbar c \alpha(0)}{8\pi^2 z^5}\uv{z}
    \label{seq:far_field_casimir_force}
\end{equation}
where $\nprad$ is the nanoparticle radius and with $\alpha(0)$ derived from the electrostatic value $\epsilon(0) = 3.8$ of the permittivity of \ch{SiO2}. The well-known formula \eqnref{seq:far_field_casimir_force} can be derived analytically from \eqnref{seq:zpf_force} (neglecting the integrand term $\propto |\alpha(\omega)|^2$) by performing a Wick rotation of the integral in \eqref{seq:zpf_force} onto the imaginary axis to obtain a non-oscillating, rapidly decaying integrand and then expanding the result for $z > \lambda$ where $\lambda$ is the minimum wavelength appearing in the integral~\cite{casimir_Phys.Rev._73:4_1948, wylie_Phys.Rev.A_30:3_1984}. 
As a consistency check, we compared a full numerical evaluation of the Wick-rotated \eqref{seq:zpf_force} (translating $\alpha(\omega)$ to the imaginary axis using the Schwarz formula $\alpha(\ii\omega) = \frac{2}{\pi}\int_0^\infty \phi\Im{\alpha(\phi)}/(\phi^2 + \omega^2) \, \dd\phi, \ \omega\in\mathbb{R}$) to the approximation \eqnref{seq:far_field_casimir_force} and found excellent agreement, confirming that the Casimir force is dominated by the electrostatic value of the polarizability also in the present case.

\begin{figure}
    \includegraphics{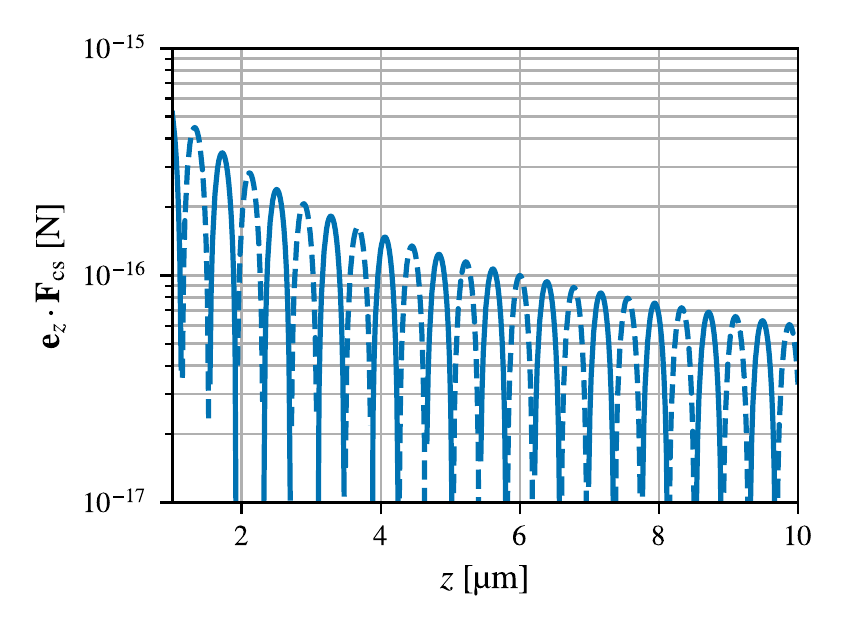}
    \caption{$\vb F_\mathrm{cs}$ for a nanoparticle above a surface which reflects the tweezer wavelength $\lambda_0$ perfectly. Solid/dashed lines indicate that the force is positive/negative in the normal direction of the surface.}
    \label{sfig:fcs}
\end{figure}

To evaluate $\vb F_\mathrm{cs}$, we approximate the optical tweezer by the time-harmonic function $\vb E_\mathrm{tw}(\vb r, t) = E_\mathrm{tw}(\vb r)\uv x\cos{(\omega_0 t)}$, where we assume the tweezer beam to be polarized parallel to the plane, consistent with the geometry of our proposed setup. This enables us to remove the integral in \eqref{seq:cs_force} to find
\begin{equation}
    \vb F_\mathrm{cs} = \frac{|\alpha(\omega_0)|^2}{2}|E_\mathrm{tw}(\vb r_0)|^2\Re{\lim_{\gpv\to\dpv}\nabla_\gpv G_{1,xx}(\vb r, \vb r_0, \omega_0)}.
\end{equation}
The field intensity at the center of the tweezer is given in terms of the experimental parameters as $|E_\mathrm{tw}(\dpv)|^2 = 2\pi N^2 P_\mathrm{tw} / \epsilon_0 c \lambda_0^2$.
We calculate the $\vb F_\mathrm{cs}$ that would result if the surface was perfectly reflecting at $\lambda_0 = 2\pi c/\omega_0$. The force would then become
\begin{equation}
    \vb F_\mathrm{cs} = -\uv z\left|\frac{\alpha(\lambda_0)}{\epsilon_0}\right|^2\frac{N^2 P_\mathrm{tw}}{64 c \lambda_0^2 z_0^4}j_x(2\pi z_0/\lambda_0)
\end{equation}
with $j_x(x) = \Re{\exp{(2\ii x)}(3-6\ii x-8x^2+8\ii x^3)}$. We plot this force in Figure~\ref{sfig:fcs}. However, in our scenario the surface is transparent at $\omega_0$ and the optical interaction force will be weaker. We approximate the magnitude of the optical interaction force above our non-perfectly reflecting surface as $|r_\perp(\omega_0)\vb F_\mathrm{cs}|$, where $r_\perp(\omega_0)$ is the reflection coefficient of the surface at $\omega_0$ and normal incidence, and the symbol $\vb F_\mathrm{cs}$ retains its meaning as the coherent scattering force above a perfectly reflecting plane.

We model the modulation of the surface reflectivity in time as follows. We assume that the presence of the surface gives rise to a Green's tensor which is effectively the Green's tensor above a perfectly reflecting surface in the range $(\omega_\mathrm{peak} - \Gamma, \omega_\mathrm{peak} + \Gamma)$ (where $\omega_\mathrm{peak} = 2\pi c/\lambda_\mathrm{peak}$ is the frequency of the dominant peak of $n(\omega,T)\Im{\alpha(\omega)}$ in the temperatures of interest and $\Gamma$ is the full width at half maximum (FWHM) of the peak), but where the Fresnel coefficients obtain a sinusoidal oscillation about their perfect reflector values with amplitude $\eta$ and frequency $\Omega_\mathrm{d}$: for $\omega \in (\omega_\mathrm{peak} - \Gamma, \omega_\mathrm{peak} + \Gamma)$,
\begin{equation}
    \left\{\begin{array}{l}
        r_s(\omega, k_\parallel) = -1 \to -(1 - \frac{\eta}{2}[1 - \cos{(\Omega_\mathrm{d} t)}]), \\
        r_p(\omega, k_\parallel) = 1 \to 1 - \frac{\eta}{2}[1 - \cos{(\Omega_\mathrm{d} t)}].
    \end{array}\right.
\end{equation}
Since the integral giving $\vb F_\mathrm{rad}$ is dominated by the value of the integrand in this range, this effectively gives
\begin{equation}
    \vb F_\mathrm{rad} \to (1-\frac{\eta}{2})\vb F_\mathrm{rad} + \frac{\eta}{2}\cos{(\Omega_\mathrm{d} t)}\vb F_\mathrm{rad}
    \label{seq:Frad_td}
\end{equation}
in the expression \eqref{seq:total_force}. Note that the symbol $\vb F_\mathrm{rad}$ retains its meaning as the strength of the force above a perfectly reflecting surface. $\vb F_\mathrm{env}$ also obtains a time-dependence in this way, but we neglect investigating it since $|\vb F_\mathrm{rad}| \gg |\vb F_\mathrm{env}|$ when $T > \Tenv$. We assume that $\vb F_\mathrm{zpf}$ does not obtain a time-dependence, since we have seen that it is sensitive only to the electrostatic limit of the integrand. Ideally, $\vb F_\mathrm{cs}$ would not obtain a time-dependence since the surface would be perfectly transparent at $\omega_0$ at all times. However, the modulation of the reflection coefficients at $\omega_\mathrm{peak}$ will realistically also lead to a residual modulation of the reflection coefficients at $\omega_0$, which in turn would give rise to a time-dependent force  $|r_\perp(t,\omega_0)\vb F_\mathrm{cs}|$ due to the interaction with the nanoparticle's optical image. As we will see later in Section~\ref{s:sensing}, to measure $\vb F_\mathrm{rad}$ we require it to dominate the $\Omega_\mathrm{d}$ frequency component of the total time-dependent force on the nanoparticle. We make this condition quantitative by defining the amplitude of the $\Omega_\mathrm{d}$ frequency component of the driving of $|r_\perp(\omega_0)|$ as $\eta_0 \equiv \lim_{T\to\infty} T^{-1} \int_{-T}^T \dd t \, \cos{(\drivingfreq t)}|r_\perp(t,\omega_0)|$. This gives the condition $|\vb F_\mathrm{cs}|\eta_0 < |\vb F_\mathrm{rad}|\eta$ where roughly $|\vb F_\mathrm{rad}|/|\vb F_\mathrm{cs}| \sim 10^{-4}$. We discuss the experimental feasibility of this requirement in Section~\ref{s:metasurface}. If the driving of the reflection coefficients at thermal frequencies is not perfectly sinusoidal, one should consider the $\eta$ appearing in this condition to be defined in an equivalent manner to $\eta_0$.
Assuming the condition on $\eta$ and $\eta_0$ to hold, the relevant effect of the time-dependent reflection coefficients on the total force $\vb F$ is
\begin{equation}
    \vb F \to \vb F - \frac{\eta}{2}\vb F_\mathrm{rad} + \frac{\eta}{2}\cos{(\Omega_\mathrm{d} t)}\vb F_\mathrm{rad}.
    \label{seq:time_dependent_total_force}
\end{equation}
To simplify the presentation in the main text, we defined the reflection coefficient $R(t,\lambda)$ to be the magnitude of the Fresnel reflection coefficient at normal incidence: $R(t,\lambda) = |r_s(2\pi c/\lambda,0;t)| = |r_p(2\pi c/\lambda,0;t)| = |r_\perp(2\pi c/\lambda,0;t)|$. We remark that the reflectivity of the surface at normal incidence is given by $R^2(t,\lambda)$.

%
%
\section{\label{s:metasurface} Experimental feasibility of the reflection modulation}

Here we discuss the experimental feasibility of modulating the surface's reflection coefficient at the peak emission wavelength $\lambda_\mathrm{peak} = \SI{8.9}{\micro m}$ with amplitude $\eta$ while keeping the modulation amplitude $\eta_0$ of the reflection coefficient at the tweezer wavelength $\lambda_0 = \SI{1.550}{\micro m}$ within the bound $\eta_0/\eta < 10^{-4}$.

The surface proposed in Figure 1(a) can be seen as an example of a dynamically tunable infrared bandstop filter operating in transmission mode \cite{wang_Mater.Res.Express_5:4_2018, wang_IEEEPhotonicsJ._12:4_2020}. An infrared bandstop filter in transmission mode is a filter which transmits most wavelengths perfectly, but (through a combination of reflection and absorption) blocks the transmission of wavelengths in a narrow band around a central infrared wavelength. Dynamical tunability means that the properties of the filter can be controlled in real time.

The realization of such filters in the form of thin materials is one of the objectives of research on metasurfaces, and several possible designs of optical bandstop filters have been reported in the literature \cite{huang_Rep.Prog.Phys._83:12_2020, kang_Adv.Opt.Mater._7:14_2019}. Examples of dynamically tunable filters in transmission mode operating in the relevant wavelength range for our proposal can be found in \cite{ju_NatureNanotech_6:10_2011, komar_Appl.Phys.Lett._110:7_2017, shi_Nanomaterials_10:12_2020}. We note that these use a variety of mechanisms, and that there is a large space of possible designs for such filter metasurfaces.

In our case, we require a filter that can be tuned between high and low reflectivity in a stopped band centered at $\lambda_\mathrm{peak}$, while maintaining a high, unmodulated transmission at the tweezer wavelength, consistent with our criterion $\eta_0/\eta < 10^{-4}$. One example of a possible metasurface design that can achieve this is a single layer of silica glass with thickness $D$ with a graphene structure on top, similar to the design displayed in Figure 1 of \cite{wang_Mater.Res.Express_5:4_2018}. The glass is transmissive to the tweezer wavelength $\lambda_0 = \SI{1.550}{\micro m}$, but conductive and therefore reflective at the thermal wavelength $\lambda_\mathrm{peak} = \SI{8.9}{\micro m}$. The graphene on the top interface modifies the standard Fabry--Perot cavity scenario posed by the single glass layer by adding conductivity $\sigma(\omega)$ to the top interface \cite{yao_NanoLett._14:11_2014}. This conductivity can be dynamically controlled by tuning the Fermi energy of the graphene structure through electric gating. By utilizing plasmonic resonances to enhance the graphene conductivity (methods to do this include (1) adding nanoantenna arrays on top of the graphene \cite{yao_NanoLett._14:11_2014, zeng_LightSciAppl_7:1_2018} and (2) patterning the graphene into ribbons \cite{ju_NatureNanotech_6:10_2011, jadidi_NanoLett._16:4_2016, wang_Mater.Res.Express_5:4_2018, wang_IEEEPhotonicsJ._12:4_2020}), the optical response of the surface at $\lambda_\mathrm{peak}$ can be dynamically switched from high reflectance to high absorption, achieving the sought modulation of the reflection coefficient in time. By careful design of the metasurface, the change in optical response at the tweezer wavelength can simultaneously be kept within the bound $\eta_0/\eta < 10^{-4}$.

To support this conclusion quantitatively, we have estimated $\eta_0/\eta$ for the metasurface proposed in Figure 1 of \cite{wang_Mater.Res.Express_5:4_2018}, which consists of a silica glass slab with a graphene ribbon structure on the top interface. In Figure 4(c) of the same reference it is predicted that, with their specific choice of ribbon geometry, a modulation of the graphene ribbon plasmon resonance suitable for the purposes of our proposal can be achieved by modulating the Fermi energy of the graphene between \SI{0.43}{eV} and \SI{0.47}{eV}. The optical tweezer field can be decoupled from the surface plasmon resonance by polarizing the electric field of the tweezer along the graphene ribbons, in which case the tweezer field interacts with the graphene ribbons as if they were a homogeneous layer of graphene \cite{ju_NatureNanotech_6:10_2011, jadidi_NanoLett._16:4_2016}. We can then estimate $\eta_0$ in this scenario by calculating the modulation of the modulus of the reflection coefficient at the tweezer wavelength due to the modulation of the Fermi energy between the above values for a metasurface consisting of a homogeneous graphene layer on top of a silica glass layer of thickness $D$  \cite{yao_NanoLett._14:11_2014}. Our estimate indicates that $\eta_0/\eta < 10^{-4}$ can be achieved by choosing the glass thickness $D$ so that the sensitivity of the modulus of the reflection coefficient at the tweezer wavelength to changes in $\sigma(2\pi c/\lambda_0)$ is minimized. The required precision in $D$ is on the order of \SI{10}{nm}, which is within the nanofabrication capabilities demonstrated in e.g.~\cite{zeng_LightSciAppl_7:1_2018}. We stress that this example is only intended to support the experimental feasibility of our proposal, and that different and/or improved designs of an optical filter that  fulfill the requirements posed by our experimental proposal can be conceived \cite{komar_Appl.Phys.Lett._110:7_2017, shi_Nanomaterials_10:12_2020}.

%
%
\section{Nanoparticle polarizability and its Lorentzian approximation}

We obtain $\alpha(\lambda)$ from measurements of the relative permittivity $\epsilon(\lambda)$ of fused silica glass in the bulk summarized in~\cite{kitamura_Appl.Opt._46:33_2007}. $\alpha(\lambda)$ is calculated from these data via the quasistatic polarizability for a dielectric sphere
\begin{equation}
    \alpha(\lambda) = 3\epsilon_0 V \frac{\epsilon(\lambda) - 1}{\epsilon(\lambda) + 2}.
    \label{seq:polarizability}
\end{equation}
The model in~\cite{kitamura_Appl.Opt._46:33_2007} gives values of $\Im{\epsilon(\lambda)}$ in the optical range $\lambda \in [1,7]\ \si{\micro m}$ which are lower than the measurements reported in the optical range in the same work. However, we have checked that the values of $\epsilon(\lambda)$ in the optical range are so small that including them in our model for $\alpha(\lambda)$ does not affect the results for the forces. Rather, we observe that the integrands giving the values of the forces are dominated by $\alpha(\lambda)$ at its peak values; the value of $\vb F_\mathrm{rad}$ is especially dominated by the peak of $\epsilon(\lambda)$ around $\lambda_\mathrm{peak} = \SI{8.9}{\micro m}$. To confirm this, we fit a frequency-domain Lorentzian model
\begin{equation}
    \alpha_L(\omega) = 3 \epsilon_0 V \frac{\omega_p^2}{\omega_\mathrm{peak}^2 - \omega^2 - \ii \Gamma \omega},
    \label{seq:lorentzian_pol}
\end{equation}
to this peak in the frequency-domain version of \eqnref{seq:polarizability}. Here $\omega_\mathrm{peak}, \omega_p$ and $\Gamma$ are all real parameters. Specifically, we fit $\Im{\alpha_L(\omega)}$ to $\Im{\alpha(\omega)}$ using the \texttt{scipy.optimize.curve\_fit} method of SciPy 1.8.0~\cite{scipy1.0}, obtaining the peak frequency $\omega_0 = 2\pi \times \SI{3.4e13}{\hertz}$, the linewidth $\Gamma = 2\pi\times\SI{1.5e12}{\hertz}$ and $\omega_p = 2\pi\times\SI{8.8e12}{\hertz}$. $\Gamma$ corresponds to a linewidth $2 \pi c \Gamma / \omega_\mathrm{peak}^2 = \SI{0.4}{\micro m}$ of $\alpha_L(\lambda)$. We then use $\alpha_\mathrm{L}$ to compute $\vb F_\mathrm{rad}$ approximately and compare to the results using $\alpha$, finding good agreement, which confirms that the peak at $\omega_\mathrm{peak}$ dominates the integral for $\vb F_\mathrm{rad}$.

\begin{figure}
    \includegraphics{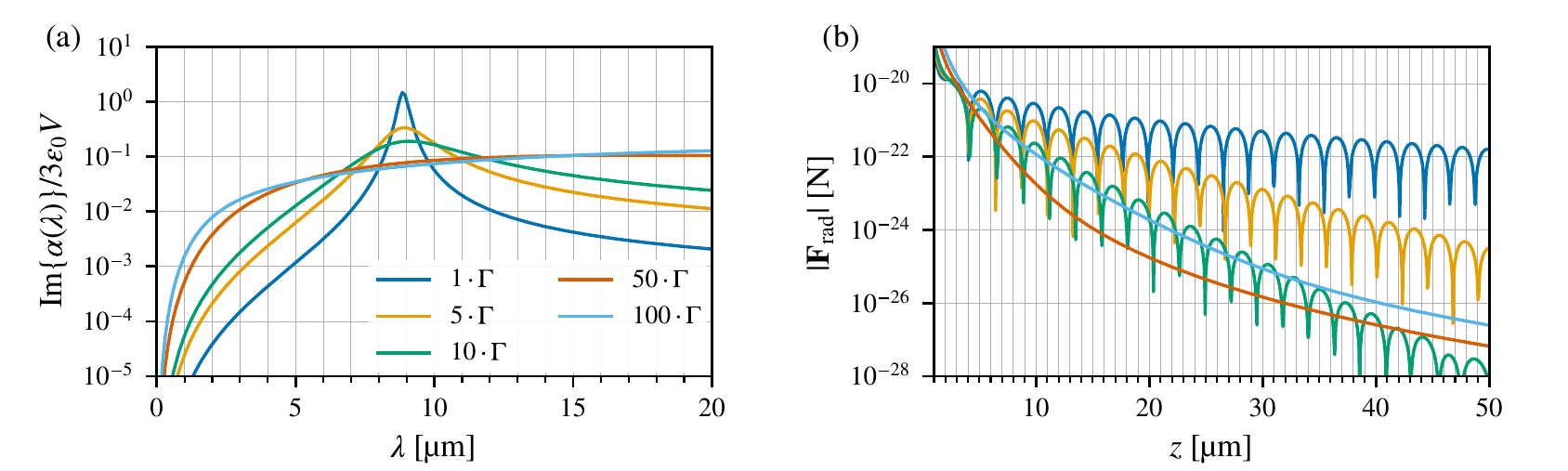}
    \caption{The effect of broadening the linewidth of the Lorentzian fit on $\vb F_\mathrm{rad}$. (a) The broadened lorentzians. The factor in the legend indicates the linewidth with respect to the original linewidth $\Gamma$. The broadened Lorentzians are renormalized in order to keep the total power radiated by the nanoparticle constant. (b) $\vb F_\mathrm{rad}$ calculated using the Lorentzian of the corresponding color in panel (a) and $T=\SI{600}{K}$.}
    \label{sfig:broadening}
\end{figure}

Using $\alpha_L$, we also investigate the effect on $\vb F_\mathrm{rad}$ of broadening the linewidth of the peak, which gives insight into the role of colored emission on the strength of $\vb F_\mathrm{rad}$. In Fig.~\ref{sfig:broadening}(a) we plot $\Im{\alpha_L(\lambda;n)}$ where $\alpha_L(\lambda;n)$ is given by \eqnref{seq:lorentzian_pol} with $\Gamma \to n\Gamma$ and renormalized to keep the total radiated power constant (see \eqnref{seq:free_space_power}), and in Fig.~\ref{sfig:broadening}(b) the $\vb F_\mathrm{rad}$ calculated using these polarizabilities. We find that as the linewidth is increased, the scaling of the force decreases progressively to $z^{-6}$, i.e.~a weaker scaling than the $z^{-5}$ far-field Casimir force, and the oscillations in sign vanish.

%
%
\section{\label{s:sensing}Sensing of $\vb F_\mathrm{rad}$}
\newcommand{\stof}{K} 
For low temperatures of the center of mass motion, an optically trapped nanoparticle is well-modelled as a damped harmonic oscillator. The equation of motion of a driven damped harmonic oscillator with position $x(t)$ is
\begin{equation}
    \ddot x + \gamma \dot x + \Omega^2 x = \frac{1}{m}\left[F(t) + \stof(t)\right].
\end{equation}
where $\gamma$ is the total damping rate of the nanoparticle's motion, $F(t)$ is a deterministic driving force and $\stof(t)$ is the sum of all the noise forces. By Fourier transform, we can solve this equation to be
\begin{align}
    x(\omega) &= \chi(\omega)\left[F(\omega) + \stof(\omega)\right] ,\\
   \chi(\omega) &= \frac{1}{m (\Omega^2 - \omega^2 - \ii\gamma\omega)}.
\end{align}
Introducing the motional and force PSDs
\begin{align}
    S_{xx}(\omega) & = \frac{1}{2\pi}\int_{-\infty}^{\infty} \expval{x(t)x(t+\tau)} \ee^{\ii\omega\tau}\,\dd\tau,
    \label{motional_psd_def}
    \\
    S_{F}(\omega) & = \frac{1}{2\pi}\int_{-\infty}^{\infty} \langle F(t)F(t+\tau)\rangle \ee^{\ii\omega\tau}\,\dd\tau,
    \label{F_force_psd_def}
    \\
    S_{\stof}(\omega) & = \frac{1}{2\pi}\int_{-\infty}^{\infty} \expval{\stof(t)\stof(t+\tau)} \ee^{\ii\omega\tau}\,\dd\tau,
    \label{R_force_psd_def}
\end{align}
one can show that the motional PSD can be expressed in terms of the force PSDs as
\begin{equation}
    S_{xx}(\omega) = |\chi(\omega)|^2\left[S_F(\omega) + S_\stof(\omega)\right].
\end{equation}
From this, we can define the thermally limited signal-to-noise ratio (SNR) as
\begin{equation}
    \SNR = \frac{S_F(\omega)}{S_\stof(\omega)}.
    \label{seq:SNR}
\end{equation}
The square root of $S_F(\omega)$ at $\SNR=1$ is called the force sensitivity $\fs$ and is given by
\begin{equation}
    \fs(\omega) = \sqrt{S_\stof(\omega)},
    \label{seq:force_sensitivity}
\end{equation}
and has units of \si{N/\sqrt{Hz}}.

To find the force PSD, we assume a harmonic driving force
\begin{equation}
    F(t) = A(z,T) \cos{(\Omega_\mathrm{d} t + \phi)},
\end{equation}
with $\Omega_\mathrm{d}$ the driving frequency and $\phi$ the initial phase of the driving, and where we find from \eqnref{seq:time_dependent_total_force} the amplitude $A(z,T)$ to be
\begin{equation}
    A(z,t) = \frac{\eta}{2} [\uv z \cdot \vb F_\mathrm{rad}(z,T)].
\end{equation}
From this, one calculates the force PSD
\begin{equation}
    S_F(\omega) = \frac{A^2(z,T)}{4}\left[\delta(\omega+\Omega_\mathrm{d}) + \delta(\omega - \Omega_\mathrm{d})\right].
    \label{seq:harmonic_force_PSD}
\end{equation}
A perfectly harmonic driving force is an idealization that does not exist in experiments. Any realistic approximately harmonic driving force will have some drift or error in its driving frequency $\Omega_\mathrm{d}$, which smoothens the $\delta$-functions appearing in \eqref{seq:harmonic_force_PSD}.
We can account for the uncertainty in $\Omega_\mathrm{d}$ by broadening the $\delta$-functions in \eqref{seq:harmonic_force_PSD} into Lorentzians with linewidth $\Gamma_\mathrm{d}$ which represent the broadening:
\begin{equation}
    \delta(\omega \pm \Omega_\mathrm{d}) \to \frac{1}{\pi}\frac{\Gamma_\mathrm{d}}{(\omega \pm \Omega_\mathrm{d})^2 + \Gamma_\mathrm{d}^2}.
\end{equation}
We can check the validity of this broadening by taking the $\Gamma_\mathrm{d}\to 0$ limit of the RHS using the Plemejl formula, and seeing that it recovers the LHS. This results in the PSD
\begin{equation}
    S_F(\omega) = \frac{A^2(z,T)}{4\pi}\left[\frac{\Gamma_\mathrm{d}}{(\omega + \Omega_\mathrm{d})^2 + \Gamma_\mathrm{d}^2} + \frac{\Gamma_\mathrm{d}}{(\omega - \Omega_\mathrm{d})^2 + \Gamma_\mathrm{d}^2}\right].
    \label{seq:pseudoharmonic_force_PSD}
\end{equation}

The contributions to the noise force $\stof(t)$ that we model are random collisions with the gas and photon shot noise~\cite{gonzalez-ballestero_Phys.Rev.A_100:1_2019, jain_Phys.Rev.Lett._116:24_2016}. 
 Thus,
\begin{gather}
    S_\stof(\omega) = S_\mathrm{photon}(\omega) + S_\mathrm{gas}(\omega), \label{seq:noise_PSD} \\
    S_\mathrm{photon}(\omega) = \frac{2}{5} \frac{\hbar\omega_0}{2\pi c^2} P_\mathrm{scatt}, \label{seq:photon_PSD} \\
    S_\mathrm{gas}(\omega) = \frac{m\gamma_\mathrm{gas}\boltzmann T_\mathrm{gas}}{\pi}. \label{seq:gas_PSD}
\end{gather}
Here $\omega_0$ is the frequency of the trapping laser, $P_\mathrm{scatt}$ is the power scattered from the tweezer laser, $\gamma_\mathrm{gas}$ is the damping rate due to gas collisions, and $T_\mathrm{gas}$ is the temperature of the gas. The power scattered from the tweezer laser is given by $P_\mathrm{scatt} = \sigma_\mathrm{scatt}I_0$, where $\sigma_\mathrm{scatt}$ is the electromagnetic scattering cross-section of the nanoparticle, and $I_0$ is the intensity at the trapping laser focus. These two quantities are given by~\cite{jain_Phys.Rev.Lett._116:24_2016}
\begin{gather}
    I_0 = \frac{P_\mathrm{tw} \omega_0^2 N^2}{2\pi c^2}, \\
    \sigma_\mathrm{scatt} = \frac{\omega_0^4|\alpha(\omega_0)|^2}{6\pi c^4 \epsilon_0^2}.
\end{gather}
Here $N$ is the numerical aperture of the trap. The gas damping rate is given by~\cite{beresnev_MotionSpherical_1990}
\begin{equation}
    \gamma_\mathrm{gas} = 0.619\frac{6\pi \nprad^2}{m} p \sqrt{\frac{2 m_\mathrm{gas}}{\pi \boltzmann T_\mathrm{gas}}}.
\end{equation}
Here $\nprad$ is the nanoparticle radius, $p$ is the gas pressure, $m_\mathrm{gas}$ is the gas molecular mass.
Inserting our expressions for the noise PSDs into the expression for $\fs$, we obtain
\begin{equation}
    \fs = \sqrt{\frac{\hbar \omega_0^7 |\alpha(\omega_0)|^2 N^2 P_\mathrm{tw}}{60\pi^3\epsilon_0^2 c^8} + 6\cdot0.619 \nprad^2 p \sqrt{\frac{2 m_\mathrm{gas} \boltzmann T_\mathrm{gas}}{\pi}}}.
\end{equation}
Putting everything together, we arrive at the expression for the signal-to-noise ratio evaluated at the driving frequency $\Omega_\mathrm{d}$
\begin{equation}
    \begin{aligned}[b]
        \mathrm{SNR}(\Omega_\mathrm{d}) & = \frac{A^2(z,T)}{4\pi\drivinglw}\left(\frac{\hbar \omega_0^7 |\alpha(\omega_0)|^2 N^2 P_\mathrm{tw}}{60\pi^3\epsilon_0^2 c^8} + 6\cdot0.619 \nprad^2 p \sqrt{\frac{2 m_\mathrm{gas}  \boltzmann T_\mathrm{gas}}{\pi}}\right)^{-1} \\
        & \approx \frac{\eta^2 [\uv z \cdot \vb F_\mathrm{rad}(z,T)]^2}{16\pi\drivinglw}\left(\frac{\hbar \omega_0^5 |\alpha(\omega_0)|^2 I_0}{30\pi^2\epsilon_0^2 c^6}\right)^{-1}
    \end{aligned}
\end{equation}
where in the second row we assumed that we are below the pressure at which $\fs$ becomes photon shot noise-dominated.
This expression lets us identify the scaling of the $\SNR$ with the experimental parameters. In this analysis, we assume a constant $T$ independently controllable from the other parameters. Since $|\vb F_\mathrm{rad}| \propto |\alpha| \propto V$, the $\SNR$ is independent of the particle size. Increasing the laser intensity $I_0$ at the nanoparticle decreases the SNR since the photon shot noise increases.

%
%
\section{\label{s:heat_balance} Heat balance equation for the nanoparticle}
\begin{figure}
    \centering
    \includegraphics{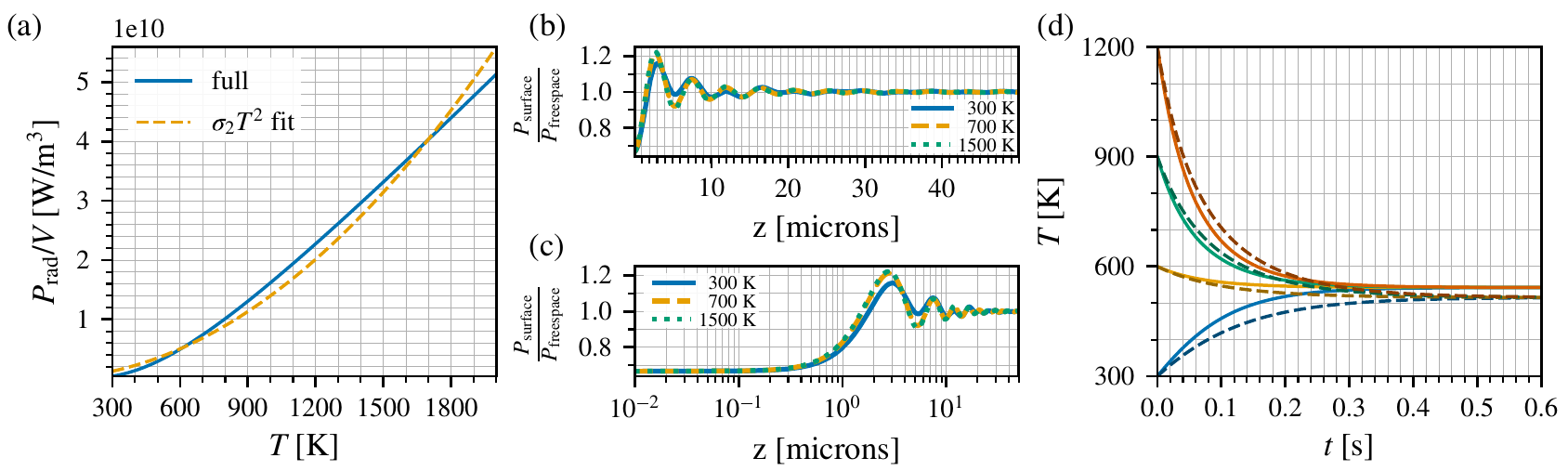}
    \caption{(a) The power radiated per volume by an \ch{SiO2} nanoparticle in free space in the temperature range relevant for experiments, as well as a quadratic fit from 300 to \SI{2000}{K} which we use to qualitatively understand the nanoparticle thermodynamics. (b) The power emitted above a perfectly reflecting surface as a fraction of the power emitted in free space. (c) The same function as in panel (b), but with a log-scaled horizontal axis. (d) Time evolution of the temperature in the radiative cooling regime for the parameters in the main text. Colors denote different initial temperatures. Solid lines are numerical solutions of \eqnref{seq:heat_balance_eq}, dashed lines are the approximation \eqnref{seq:approx_T_dyns}. The numerical steady state temperature is \SI{543}{K} while the approximation gives $T_\infty = \SI{515}{K}$.}
    \label{sfig:power}
\end{figure}
Our goal in this section is to capture the behavior of the temperature dynamics predicted by fluctuational electrodynamics in a simple, qualitative model. We predict the temperature dynamics of the nanoparticle for our experimental parameters using the heat balance equation
\begin{equation}
    c_m \rho V\dfdx{T}{t} = P_\mathrm{abs} - P_\mathrm{rad}(T),
    \label{seq:heat_balance_eq}
\end{equation}
where we use $c_m = \SI{700}{J/kg}$ for \ch{SiO2}, and
\begin{equation}
    P_\mathrm{abs} = P_\mathrm{abs,tw} + P_\mathrm{abs,env}(\Tenv),
\end{equation}
Here $P_\mathrm{abs,env}(\Tenv)$ is the power that the nanoparticle absorbs from environmental radiation, and $P_\mathrm{abs,tw}$ is the power that the nanoparticle absorbs from the laser beam. The latter is
\begin{equation}
    P_\mathrm{abs,tw} = \frac{\omega_0\Im{\alpha(\omega_0)}}{c} I_0.
    \label{seq:absorbed_laser_power} 
\end{equation}
As can be calculated using $P_\mathrm{rad}(T) - P_\mathrm{abs,env}(\Tenv) = \expval{[\partial_t\vb d(t)] \cdot \vb E(t)}$ (see e.g.~\cite{messina_Phys.Rev.B_88:10_2013}),
 in free space
\begin{equation}
    P_\mathrm{rad}(T) = \frac{\hbar c^2}{\pi^2}\int_0^\infty\dd k\, \frac{k^4}{e^{\hbar c k/k_\mathrm{B} T} - 1}\Im{\alpha(ck)},
    \label{seq:free_space_power}
\end{equation}
and $P_\mathrm{abs,env}(\Tenv) = P_\mathrm{rad}(\Tenv)$. We have not included the correction to \eqnref{seq:free_space_power} from radiation reaction (see \cite{rubiolopez_Phys.Rev.B_98:15_2018} and references therein) because it gives negligible corrections for the material and particle size that we consider. With our parameters, $P_\mathrm{abs,env}(\Tenv) \ll P_\mathrm{abs,tw}$, so we neglect $P_\mathrm{abs,env}(\Tenv)$. We obtain a simplified model of $P_\mathrm{rad}(T)$ for the silica nanoparticle by approximating it as a quadratic function of $T$:
\begin{equation}
    P_\mathrm{rad}(T) \approx \sigma_2 V T^2,
    \label{seq:quadratic_power}
\end{equation}
where we find $\sigma_2 = \SI{2e4}{W/m^3K^2}$ by numerical fit to~\eqref{seq:free_space_power}. We display~\eqnref{seq:free_space_power} and compare it to the quadratic approximation~\eqnref{seq:quadratic_power} in Fig.~\ref{sfig:power}(a). $P_\mathrm{rad}(T)$ is in general affected by the presence of dielectric material like the surface. As an example, we plot $P_\mathrm{rad}(T)$ for our silica nanoparticle above a perfectly reflecting surface in Fig.~\ref{sfig:power}(b) and (c). The emitted power oscillates with $z$ with a maximum change of \SI{20}{\percent} for $z>\SI{1}{\micro m}$. The qualitative behavior of the temperature dynamics will not be affected by this change, so we find that \eqnref{seq:quadratic_power} still qualitatively captures the emitted power for this example. To make a qualitative model of the temperature dynamics, we insert \eqnref{seq:quadratic_power} into \eqref{seq:heat_balance_eq}. This gives a simple instance of Ricatti's equation for $T$ with solution
\begin{equation}
    T(t) = \left\{\begin{matrix} T_\infty\tanh{[t/\tau + \mathrm{artanh}\,(T_0/T_\infty)]} & (T_\infty > T_0) \\
    T_\infty\coth{[t/\tau + \mathrm{artanh}\,(T_\infty/T_0)]} & (T_\infty < T_0) \end{matrix}\right.
   \label{seq:approx_T_dyns}
\end{equation}
where the time constant is
\begin{equation}
    \tau = \frac{\rho c_m}{\sqrt{\frac{P_\mathrm{abs}}{V}\sigma_2}},
\end{equation}
and $T_\infty = \sqrt{P_\mathrm{abs}/V \sigma_2}$. Note that $P_\mathrm{abs}/V$ is independent of $V$. In Fig.~\ref{sfig:power}(d) we compare time trajectories of the temperature obtained with numerical solution of \eqnref{seq:heat_balance_eq} and with the approximation \eqnref{seq:approx_T_dyns}. We find that the approximation reproduces the qualitative behavior of the trajectories and that the final temperature agrees within \SI{30}{K}.

\end{document}